\iftrue
\documentclass[aps,prl,twocolumn,
			   groupedaddress,superscriptaddress,
			   amsfonts,amssymb,amsmath,
			   citeautoscript,
			   a4paper]{revtex4-1}
\else
\documentclass[aps,pra,preprint,
			   groupedaddress,superscriptaddress,notitlepage,
			   amsfonts,amssymb,amsmath,
			   citeautoscript,
			   a4paper]{revtex4-1}
\fi

\usepackage[pdftex]{hyperref}
\hypersetup{colorlinks,
			linkcolor={blue!75!black!80!yellow},
			citecolor={blue!75!black!80!yellow},
			urlcolor={blue!75!black!80!yellow},
			pdfstartview=FitH}
\usepackage{graphicx}
\usepackage{mathrsfs}
\usepackage{amsthm}
\usepackage{physics}
\usepackage{xspace}
\usepackage{braket}
\usepackage{xr}
\usepackage{gensymb} 
\usepackage{xcolor,soul}
\usepackage{stmaryrd} 
\usepackage[UKenglish]{babel}
\usepackage{placeins} 
\usepackage{siunitx}
\DeclareSIUnit[number-unit-product=]\percent{\char`\%} 
\usepackage{makecell}
\usepackage{diagbox}
\usepackage{float}
\usepackage{amsmath} 
\usepackage{empheq} 

\usepackage{txfonts}  
\usepackage{txfontsb} 

\newcommand*\diff{\mathop{}\mathrm{d}}

\renewcommand{\Re}{\operatorname{Re}}
\renewcommand{\Im}{\operatorname{Im}}
\newcommand{\iu}{\mathrm{i}}
\newcommand{\e}{\mathrm{e}}

\newcommand{\appropto}{\mathrel{\vcenter{
			\offinterlineskip\halign{\hfil$##$\cr
				\propto\cr\noalign{\kern2pt}\sim\cr\noalign{\kern-2pt}}}}}

\newcommand{\ie}{i.e.\@\xspace}  

\newcommand{\eg}{e.g.\@\xspace}

\newcommand{\thL}{\theta_{\mathrm{L}}}
\newcommand{\thR}{\theta_{\mathrm{R}}}
\newcommand{\JL}{J_{\mathrm{L}}}
\newcommand{\JR}{J_{\mathrm{R}}}
\newcommand{\nL}{n_{\mathrm{L}}}
\newcommand{\nR}{n_{\mathrm{R}}}
\newcommand{\nE}{n_{\mathrm{E}}}

\makeatletter
\newcommand*{\addFileDependency}[1]{
  \typeout{(#1)}
  \@addtofilelist{#1}
  \IfFileExists{#1}{}{\typeout{No file #1.}}
}
\makeatother

\usepackage{scalerel}

\usepackage{textcomp} 
\usepackage{xifthen}
\usepackage{etoolbox}
\newboolean{togglecomments}
\newboolean{togglechanges} 

\setboolean{togglecomments}{true}
\setboolean{togglechanges}{false}

\newcommand{\comment}[2]{%
    \ifbool{togglecomments}%
    {\textcolor{blue!70!black}{\small\textsf{%
    \textsuperscript{\textsc{\textsf{\MakeLowercase{#1}}}}%
    [#2]}}} 
    {}}     
\newcommand{\swap}[2]{\ifbool{togglechanges}
    {#2}  
    {\textcolor{red!70!black}{[#1]}\textrightarrow{}\textcolor{green!50!black}{[#2]}}}
\newcommand{\remove}[1]{\ifbool{togglechanges}
    {}    
    {\textcolor{red!70!black}{#1}}}
\newcommand{\inset}[1]{\ifbool{togglechanges}
    {#1}  
    {\textcolor{green!50!black}{#1}}}
\newcommand{\optional}[1]{\ifbool{togglechanges}
    {#1}  
    {\textcolor{yellow!50!orange!80!gray}{#1}}}
\newcommand{\citeremind}[1]{%
    [\textcolor{blue!75!black!80!yellow}{
        $\blacksquare$%
           \ifthenelse{\isempty{#1}}
               {}
               {\textsuperscript{\textsf{#1}}}%
        }]\xspace}
\newcommand{\todo}[1]{
    \textcolor{orange!80!yellow!95!black}{\textbf{[}%
        \ifthenelse{\isempty{#1}}%
        {\text{$\blacksquare$}}%
        {{\small\textsf{#1}}}%
        \textbf{]}}}
        
\newcommand{\hkuaffil}{\footnotesize Department of Physics, The University of Hong Kong, Pokfulam, Hong Kong, China}
\newcommand{\sstaffil}{\footnotesize College of Optical-Electrical Information and Computer Engineering,
University of Shanghai for Science and Technology, Shanghai 200093, China}

\begin{document}

\title{Synthetic non-Abelian gauge fields for non-Hermitian systems}

\author{Zehai Pang}
\affiliation{\hkuaffil}
\author{Jinbing Hu}
\affiliation{\hkuaffil}
\affiliation{\sstaffil}
\author{Yi~Yang}
\email{yiyg@hku.hk}
\affiliation{\hkuaffil}

\begin{abstract}
Non-Abelian gauge fields are versatile tools for synthesizing topological phenomena but have so far been mostly studied in Hermitian systems, where gauge flux has to be defined from a closed loop in order for gauge fields, whether Abelian or non-Abelian, to become physically meaningful.
We show that this condition can be relaxed in non-Hermitian systems by proposing and studying a generalized Hatano--Nelson model with imbalanced non-Abelian hopping. Despite lacking gauge flux in one dimension, non-Abelian gauge fields create rich non-Hermitian topological consequences.
Under only nearest-neighbor coupling, non-Abelian gauge fields enable Hopf-link bulk braiding topology, whose phase transition accompanies the emergence of exceptional points (EPs).
At both ends of an open chain, non-Abelian gauge fields lead to the simultaneous presence of non-Hermitian skin modes, whose population can be effectively tuned.
Asymptotic analysis shows that this tuning mechanism stems from the interplay between the Abelian Hatano--Nelson coupling and effective high-order hopping, which becomes substantial near the EP phase transition condition. 
The predicted non-Hermitian phenomena, enabled by non-Abelian gauge fields, could be realized in synthetic dimensional optical platforms such as time-multiplexed photonic mesh lattices and driven ring resonators.
\end{abstract}

\maketitle


%
\indent Open physical systems interacting with external environments are described by non-Hermitian Hamiltonians that support complex eigenvalues. Compared to closed systems, non-Hermitian systems exhibit rich unique phenomena, such as power oscillations~\cite{musslimani2008optical,makris2008beam,regensburger2012parity}, unidirectional invisibility~\cite{feng2013experimental,regensburger2012parity}, and exceptional-point (EP) encirclement~\cite{gao2015observation,doppler2016dynamically},
which have no counterparts in Hermitian systems. 
Besides their bulk invariants defined from eigenvectors~\cite{yao2018edge,song2019non,kunst2018biorthogonal,xiong2018does} as in Hermitian systems, non-Hermitian systems also exhibit eigenvalue topology~\cite{bergholtz2021exceptional,gong2018topological,kawabata2019symmetry,yokomizo2019non,zhang2020correspondence,okuma2020topological,yang2020non,wang2021topological,hu2021knots,borgnia2020non} due to the expansion of eigenenergies from the real to the complex regime. 
Importantly, non-Hermitian eigenstates of a non-vanishing eigenvalue winding number are all localized at the end of open systems known as the non-Hermitian skin effect (NHSE)~\cite{lee2016anomalous,yao2018edge,xiong2018does}. NHSE has been implemented widely in photonics~\cite{weidemann2020topological,wang2021generating,wang2021topological,xiao2020non,liu2022complex}, acoustics~\cite{zhang2021observation,zhang2021acoustic,gu2022transient}, mechanics~\cite{wang2022non,ghatak2020observation,scheibner2020non}, and electric circuits~\cite{li2020critical,helbig2020generalized,hofmann2020reciprocal,liu2021non,zou2021observation,shang2022experimental}.
Moreover, synthetic gauge fields have been introduced for better controlling non-Hermitian systems~\cite{lin2021steering,midya2018non,longhi2017non,longhi2015non,longhi2017nonadiabatic,wong2021topological}, but most efforts have been dedicated to Abelian gauge fields.

Non-Abelian physics has recently attracted lots of attention in acoustics and photonics~\cite{chen2019non,yang2019synthesis,yang2020observation,guo2021experimental,jiang2021experimental,brosco2021non,chen2022classical,sun2022non,you2022observation,noh2020braiding,zhang2022non,xu2016simulating,yang20202d,yang2022non,gianfrate2020direct,whittaker2021optical,iadecola2016non,polimeno2021experimental,cheng2023artificial}. In particular, non-Abelian gauge fields, leveraging the internal degrees of freedom of particles, are a synthetic control knob for realizing non-Abelian physics in engineered physical systems~\cite{aidelsburger2018artificial}.
These gauge fields enable synthetic spin-orbit interaction and can be used for creating non-Abelian Aharonov--Bohm interference and lattice models featuring complex gauge structures.
Moreover, recent experiments have demonstrated the possibility of creating and tuning building blocks of non-Abelian gauge fields in fibers~\cite{yang2019synthesis} and circuits~\cite{wu2022non}, indicating their applicability for large lattice systems.
The effectiveness of synthetic gauge fields substantially relies on their dimensionality.
In particular, pure one-dimensional (1D) systems forbid the definition of closed loops and the associated magnetic flux.
Thus, synthetic gauge fields, whether Abelian or non-Abelian, carry little physical consequences in 1D Hermitian systems.
Although the 1D spin-orbit interaction realized with cold atoms~\cite{lin2011spin} seems to be a counterexample, an extra Zeeman term has to be added for the Rashba--Dresselhaus gauge fields to become nontrivial.
So far, non-Abelian gauge fields have seldom been explored in non-Hermitian systems, where the dimensionality constraint above could be violated.

\begin{figure*}
    \centering
    \includegraphics[width=\linewidth]{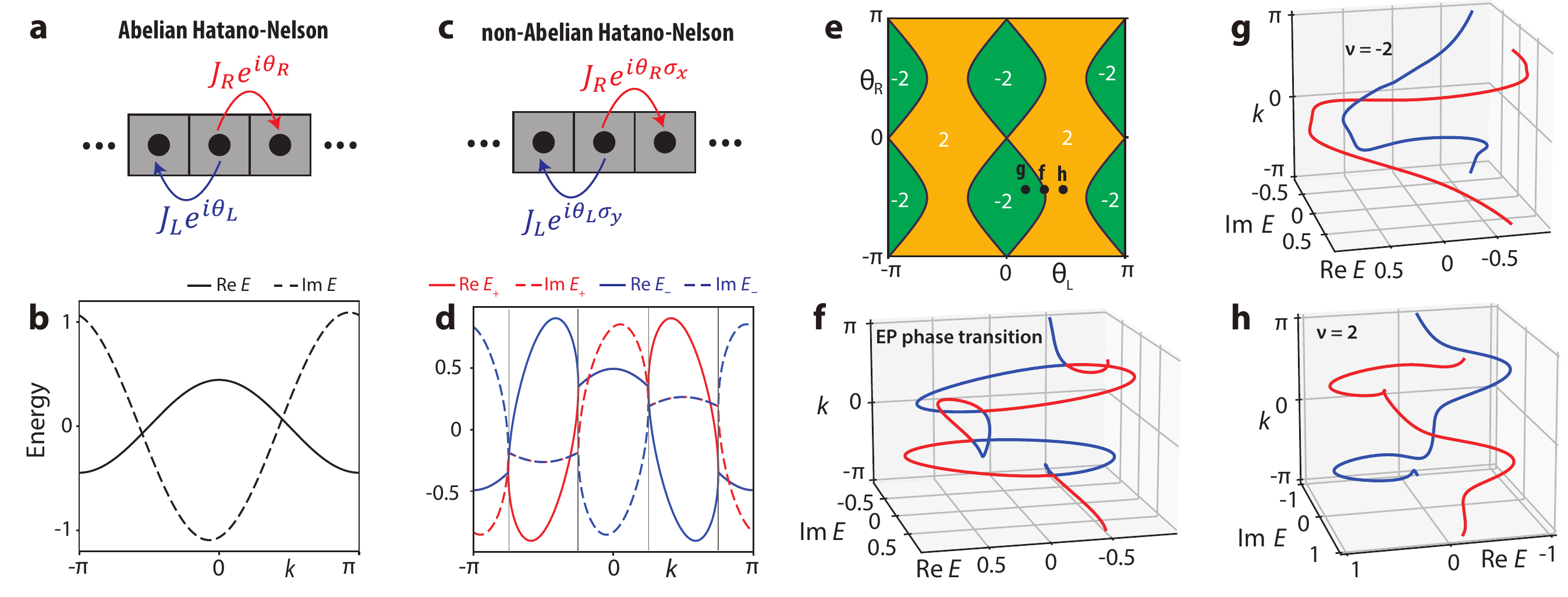}
    \centering
    \caption{\textbf{Hatano--Nelson model with Abelian and non-Abelian gauge fields.}
             \textbf{a-b.} Hatano--Nelson model (a) with nonreciprocal U(1) hopping phases and its single non-Hermitian band (b) exhibiting a winding number $w = 1$.
             \textbf{c.} Hatano--Nelson model with nonreciprocal SU(2) hopping phases.
             \textbf{d.} Two non-Hermitian bands ($E_+$ and $E_-$) exhibiting four exceptional points (EPs) at momenta $\left\{\pm\pi/4,\pm3\pi/4\right\}$ (vertical grey lines). 
             \textbf{e.} Phase diagram of the non-Abelian Hatano--Nelson model under $\JL/\JR=7/6$, featuring Hopf links with braid degrees $\nu=\pm2$. Their phase transition accompanies the appearance EPs (black solid lines).
             \textbf{f-h.} Braiding of the two bands under the EP phase transition (f), $\nu=-2$ (g), and $\nu=2$ (h) Hopf-link phases in the ($\Re E$, $\Im E$, $k$) space.
             Here we use $\JL = 0.7$, $\JR = 0.6$, and $\thR = -1.38$ throughout; $\thL = -1$ for b,d, and f, $\thL = -0.5$ for g, and $\thL = -1.5$ for h, respectively, as shown by black dots in e.
    }
    \label{fig:model}
\end{figure*}

In this work, we introduce non-Abelian gauge fields to generate non-Hermiticity and study their manipulation of NHSE.
Under bare nearest-neighbor coupling, the bulk spectrum of a non-Abelian Hatano--Nelson model forms Hopf links in complex energy braiding, which was realized only with long-range hopping previously.
The Hopf links are characterized by the braiding degree $\pm 2$, whose phase transition accompanies the appearance of EPs. 
Non-Abelian gauge fields enable the simultaneous presence of left- and right-localized skin modes whose population is tunable, confirmed by both winding-number and the non-Bloch calculations.
Asymptotic analyses further reveal that the tunability is most effective near the EP phase transition, resulting from the interplay between the Abelian Hatano--Nelson coupling and effective high-order hopping on the order of $4N-1$ ($N$ is an integer). 

The Hatano--Nelson model~\cite{hatano1996localization} is a prototypical 1D system that demonstrates NHSE because of its nonreciprocal hoppings. 
We first extend the model with U(1) Abelian gauge fields as (Fig.~\ref{fig:model}a)
\begin{align}
    \hat{H}_{0} = \sum_{m} \JL c^{\dagger}_{m}\mathrm{e}^{\iu \thL}c_{m+1} + \JR c^{\dagger}_{m+1}\mathrm{e}^{\iu \thR}c_{m}.
    \label{eq:HN}
\end{align}
Here $c^{\dagger}_m (c_m)$ is the creation (annihilation) operator at site $m$, $J_\mathrm{L(R)}$ is the real hopping amplitude leftward (rightward), and $\theta_\mathrm{L(R)}$ the corresponding hopping phases.
The conventional Hatano--Nelson model is restored if $\thL=\thR=0$. One can reformulates Eq.~\eqref{eq:HN} as 
$
H_0(k)\mathrm{e}^{-\iu\theta_+} =\JL\mathrm{e}^{\iu(k+\theta_- )} + \JR\mathrm{e}^{-\iu(k+\theta_- )},
$
where $\theta_+ = \left(\thL+\thR\right)/2$ and $\theta_- = \left(\thL-\thR\right)/2$.
As in Hermitian systems, a Peierls substitution of $\theta_-$ acts on the momentum $k$.
Meanwhile, on the left-hand side, a Peierls substitution of $\theta_+$ acts on the complex energy, i.e. a rotation on the complex energy plane.
Thus, the U(1) fields only lead to trivial modifications to the Hatano--Nelson model.
This is confirmed by the energy band shown in Fig.~\ref{fig:model}b, which exhibits a winding number $w=+1$ on the complex energy plane, where $w \equiv \frac{1}{2\pi} \int_0^{2\pi}  \partial_k \mathrm{arg}(E(k)-E_\mathrm{b}) \diff k = \mathrm{sgn}(\JL-\JR)$ and $\pm1$ indicates counter-clockwise (CCW) and clockwise (CW) rotation, respectively.

In contrast, the model gets substantially modified with SU(2) non-Abelian gauge fields (Fig.~\ref{fig:model}c):
\begin{align}\label{eq:HN_nab_r}
    \hat{H} = \sum_{m} \JL c^{\dagger}_{m}\mathrm{e}^{\iu \thL\sigma_y}c_{m+1} + \JR c^{\dagger}_{m+1}\mathrm{e}^{\iu \thR\sigma_x}c_{m}, 
\end{align}
where $\sigma_x$ and $\sigma_y$ are Pauli matrices.
Notably, in Eq.~\eqref{eq:HN_nab_r}, both the hopping amplitudes $\left(\JL,\JR\right)$ and the non-Abelian hopping phases $\left(\thL,\thR\right)$ contribute to non-Hermiticity. This feature distinguishes our system from a recent study on non-Hermitian Aubry--Andr\'{e}--Harper models~\cite{zhou2023non}, where the non-Abelian on-site potentials alone do not cause non-Hermiticity.
The Bloch Hamiltonian of Eq.~\eqref{eq:HN_nab_r} is
\begin{align}\label{eq:HN_nab_k}
    H(k) &= A(k)\sigma_0 + \iu \JL\sin{\thL}\mathrm{e}^{\iu k}\sigma_y +\iu \JR\sin{\thR}\mathrm{e}^{-\iu k}\sigma_x,
\end{align}
where 
\begin{align}
    A(k) = \JL\cos{\thL}\mathrm{e}^{\iu k} + \JR\cos{\thR}\mathrm{e}^{-\iu k}
\end{align}
and $\sigma_0$ the identity matrix. The eigen-energy of $\hat{H}$ is given by
\begin{align} \label{eq:HN_nab_E}
    E_{\pm}(k) = A(k)\pm \iu\sqrt{\JL^2\sin^2{\thL}\mathrm{e}^{\iu2k} + \JR^2\sin^2{\thR}\mathrm{e}^{-\iu2k}}.
\end{align}
Eq.~\eqref{eq:HN_nab_E} permits EPs at $k_\mathrm{EP}=\left\{\pm\pi/4,\pm3\pi/4\right\}$ when 
\begin{align}
    \JL^2\sin^2{\thL} = \JR^2\sin^2{\thR}
    \label{eq:EP}
\end{align}
the EP condition is satisfied, as shown by an example spectrum in Fig.~\ref{fig:model}d. 

The two energy bands in Eq.~\eqref{eq:HN_nab_E} form Hopf link in ($\Re E$, $\Im E$, $k$) space (different from the exceptional-line links in three-dimensional momentum space~\cite{yang2019non,carlstrom2018exceptional}).
In fact, the EP condition Eq.~\eqref{eq:EP} is the phase transition of the energy braiding between two types of Hopf links, defined by a braiding degree~\cite{wang2021topological} $\nu=\pm2$ (Fig.~\ref{fig:model}e), where
\begin{align}
    \nu\equiv\int_{0}^{2\pi} \frac{\diff k}{2\pi\iu}\frac{\diff}{\diff k}\mathrm{ln}\;\mathrm{det}\left(\hat{H}_k - \frac{1}{2}\mathrm{Tr}\hat{H}_k\right).
\end{align} 
Fig.~\ref{fig:model}f-g confirm this transition, where Hopf links of opposite braiding degrees (Fig.~\ref{fig:model}g and h) appear on opposite sides of the EP phase transition (Fig.~\ref{fig:model}f).
Non-Hermitian energy braiding of the Hopf-link type has been identified previously but requires longer range hopping, such as the next-nearest-neighbor coupling~\cite{wang2021topological,hu2021knots}. In a recent paper \cite{hu2021knots}, the Hopf link is achieved by using a building block Hamiltonian $\left(0, \e^{\iu n k};1, 0\right)$, where $n=2$ gives rise to Hopf link.
Nevertheless, here non-Abelian gauge fields enable the realization of the Hopf link using nearest-neighbor coupling only.
Therefore, even though no gauge flux can be defined, introducing non-Abelian gauge fields can sufficiently drive non-Hermitian topological phase transitions in a 1D bulk, which is impossible for Hermitian systems.
\begin{figure}[htbp]
    \centering
    \includegraphics[width=\linewidth]{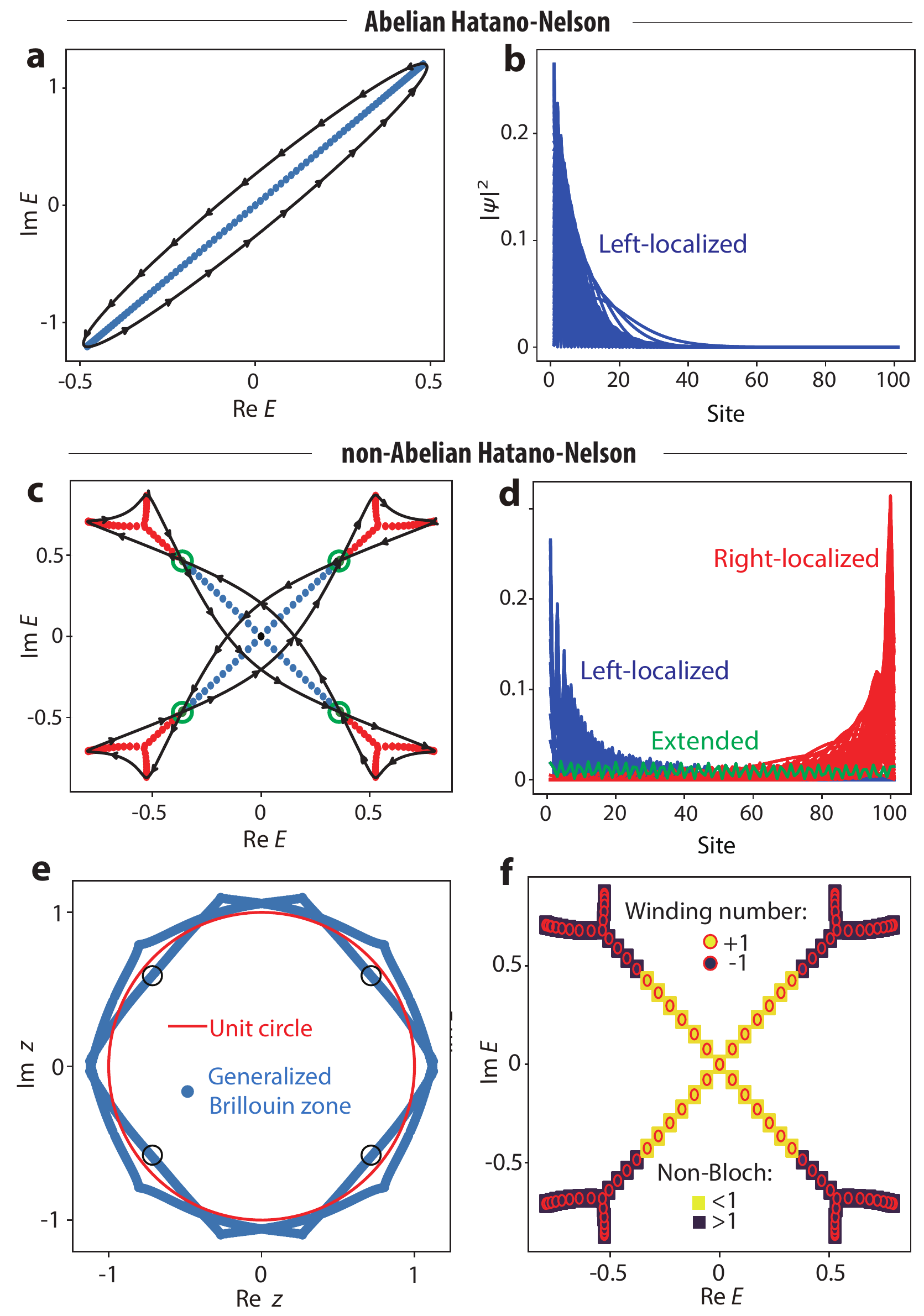}
    \caption{\textbf{Analysis of non-Hermitian skin effect.} 
    \textbf{a-d.} Periodic- (black lines) and open-boundary (dots) spectra (a and c) and eigenstates (b and d) of the Abelian (a-b) and non-Abelian (c-d) Hatano--Nelson models.
    \textbf{e.} Localization analysis. The unit circle (red) intersects the generalized Brillouin zone (blue), indicating the simultaneous presence of left- (inside the unit circle) and right-localized (outside the unit circle) states. The black circles denote the zero-mode solutions. 
    \textbf{f.} Simultaneous left- and right-localization confirmed by consistent non-Bloch (squares) and winding-number (circles) calculations. 
    Here, $\JL = 0.7$, $\JR = 0.6$, $\thL = -2.5$, $\thR = -1.4$.
    }
    \label{fig:skin}
\end{figure}

Next, we show how non-Abelian gauge fields enrich NHSE in Fig.~\ref{fig:skin}.
We calculate the eigen-spectra of $H_0$ and $H$ under the periodic boundary condition (PBC) and the open boundary condition (OBC).
Both the PBC spectra of $H_0$ and $H$ form closed loops surrounding the zero energy in the complex plane, indicating point-gapped bulk topology, while their OBC spectra become open arcs~\cite{zhang2022review, ashida2020non, zhang2020correspondence}. 
Thus, the PBC and OBC spectra are topologically distinct, and NHSE inevitably occurs in both models as a consequence of topological phase transition~\cite{okuma2020topological}.
In Fig.~\ref{fig:skin}a, a PBC spectrum of the Abelian $H_0$ is an ellipse showing uniform CCW winding, and the corresponding OBC arc occupies the major axis. 
As determined by $\abs{\JL/\JR}>1$, all OBC states are localized on the left of the chain (Fig.~\ref{fig:skin}b), identical to the conventional Hatano--Nelson model~\cite{gong2018topological}; it confirms our previous analysis that the U(1) Abelian gauge field only leads to a trivial Peierls substitution without modifying the NHSE.

In contrast, the NHSE in the non-Abelian Hatano--Nelson model $H$ is far richer.
The PBC spectrum of $H$ (Fig.~\ref{fig:skin}c), under the same set of parameters, simultaneously exhibits CW and CCW winding at the four corners and center of the Hopf link, respectively. 
Consequently, the OBC arc enclosed by these sectors should demonstrate leftward (blue in Fig.~\ref{fig:skin}c) and rightward (red in Fig.~\ref{fig:skin}c) localization, respectively, whose simultaneously presence is confirmed by the visualization of the eigenstates in Fig.~\ref{fig:skin}d.
Meanwhile, extended states (green in Fig.~\ref{fig:skin}d) also appear at the boundary of the CW and CCW winding (green circles in Fig.~\ref{fig:skin}c).
This simultaneous leftward and rightward localization cannot be explained solely by the imbalanced hopping amplitudes $(\JL,\JR)$. 

We adopt both the non-Bloch~\cite{yao2018edge, yang2020non, yokomizo2019non} and the winding-number approaches~\cite{zhang2020correspondence} to study the effect from non-Abelian gauge fields.
Using the non-Bloch approach, for an OBC energy $E$, we calculate the characteristic polynomial $\mathrm{det}[H(z)-E]$, a quartic function of $z$ (see Sec.~S2 of Ref.~\cite{SM_note}) The generalized Brillouin zone (GBZ) \cite{yokomizo2019non} $C_z$ is thus determined by the trajectory of $z_2$ and $z_3$ under the condition $|z_2| = |z_{3}|$, where $z_2$ and $z_3$ are the second and third solutions to the polynomial sorted by the absolute values in ascending order. 
In parallel, we calculate the multi-band winding number (details in Sec.~S3)
$
w \equiv \sum_{n=1}^{N} \int_{-\pi}^{\pi} \frac{\diff k}{2\pi} \partial_k \mathrm{arg}(E_n(k)-E_\mathrm{b}),
$
where $E_\mathrm{b}$ is a complex OBC energy base point, $n$ labels the band index, and $N$ is the total number of bands. 
The sign of $w$ indicates the localization, \ie $w>0$ and $w<0$ for left- and right-localization, respectively, and $w=0$ for extended states. 
Shown in Fig.~\ref{fig:skin}f, Consistency is achieved between our non-Bloch and winding-number analyses.

We specifically discuss the properties of zero modes $E = 0$ (details in Sec.~S2) under OBC. The associated all four non-Bloch solutions are $z = \pm\iu \sqrt{\JR/\JL\mathrm{e}^{\pm\iu\alpha}}$, where $\alpha = \arctan(\sqrt{1-F^2}/F)$ and $F = \cos(\thR)\cos(\thL)$. All roots of the characteristic polynomial have equal absolute values, guaranteeing the existence of OBC zero modes, as can be seen by the pinned crossing at $E=0$ in the complex plane (Fig.~\ref{fig:skin}c). We also prove that the OBC zero modes must be doubly degenerate (Sec.~S2). 
Furthermore, The absolute values of the zero-mode non-Bloch solutions depend only on the ratio of the hopping amplitudes, which indicates that even SU(2) gauge fields cannot modify the localization direction of the zero modes (proof in Sec.~S2). 

Nevertheless, the localization of non-zero modes can be effectively manipulated by non-Abelian gauge fields.
The GBZ of the non-Abelian model, shown in Fig.~\ref{fig:skin}e, exhibits states both inside and outside the unit circle; these states are thus associated with left- and right-localization, respectively.
We prove that this tunability of the skin modes is not possible using the U(1) gauge fields in the Abelian Hatano--Nelson model [see Eq.~\eqref{eq:HN} and proof in Sec.~S2.A].

\begin{figure}[htbp]
    \centering
    \includegraphics[width=\linewidth]{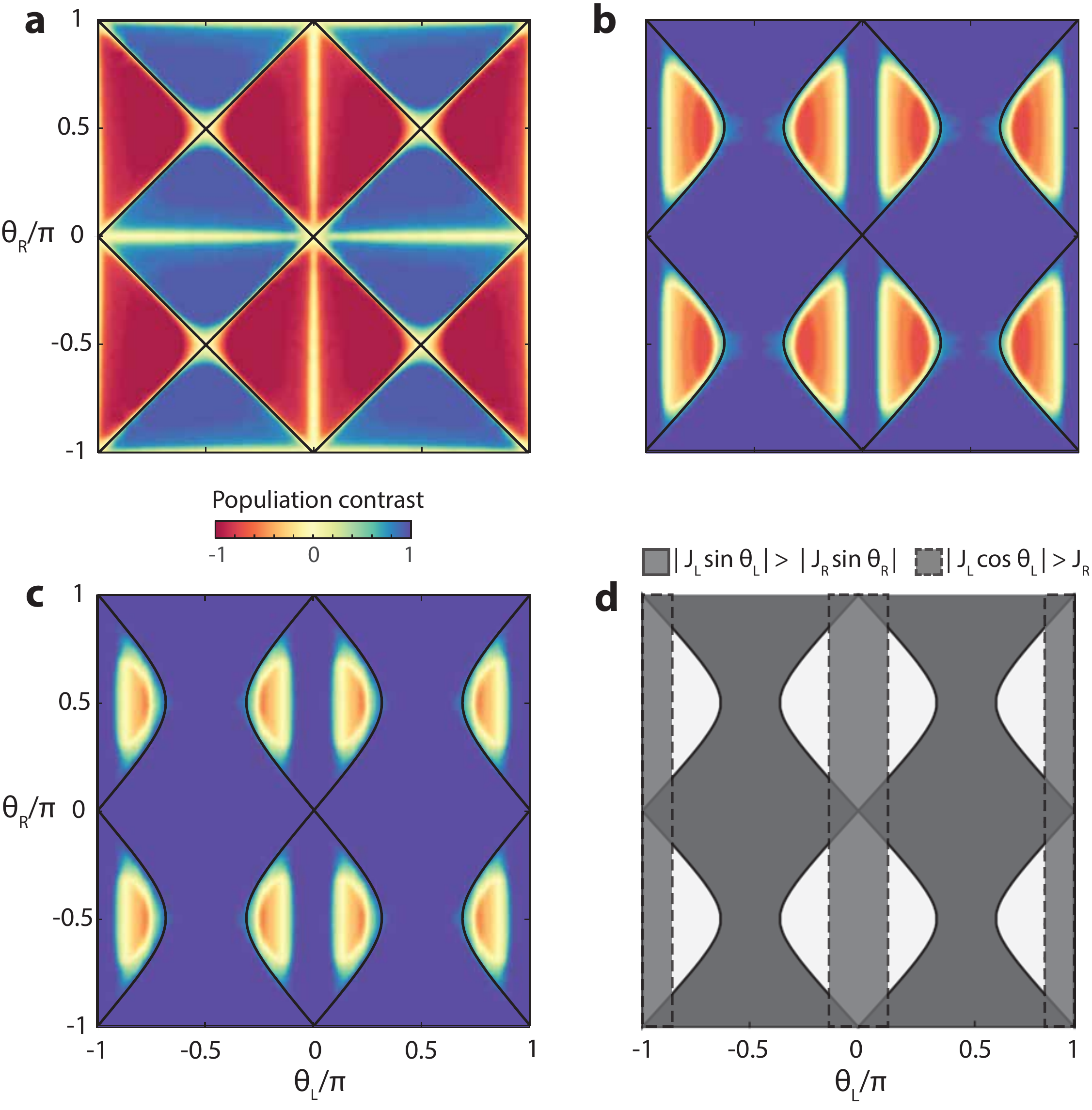}
    \centering
    \caption{\textbf{Tuning non-Hermitian localization with non-Abelian gauge fields.}
             \textbf{a-c.} Population contrast $\eta$ [Eq.~\eqref{eq:contrast}] in the $(\thL,\thR)$ space when $(\JL,\JR) = (0.5,0.5)$ (a), $(0.55,0.5)$ (b), and $(0.6,0.5)$ (c), respectively. The black solid lines define the exceptional-point phase boundary between the two braiding degrees $\nu=\pm2$ (see Fig.~\ref{fig:model}e).
             \textbf{d.} Analytical asymptotic analysis. 
             Effective high-order hoppings [Eqs.~\eqref{eq:asymp_2} and \eqref{eq:asymp_3}] are non-negligible in the vicinity of the white region, where the localization tunability becomes most effective. In d, we use $(\JL,\JR) = (0.55,0.5)$, the same as those in b.
    }
    \label{fig:contrast}
\end{figure}

To further elucidate the interplay between the imbalanced hopping amplitudes and non-Abelian gauge fields, we define a population contrast $\eta$ in the OBC eigenstates as
\begin{align}
    \eta(\JL,\JR,\thL,\thR)\equiv\frac{\nL-\nR}{\nL+\nR+\nE},
    \label{eq:contrast}
\end{align}
where $\nL$, $\nR$, and $\nE$ are the number of left-localized, right-localized, and extended states, respectively.
Fig.~\ref{fig:contrast} shows the population contrast as a function of the gauge fields $(\thL,\thR)$ under different choices of $(\JL,\JR)$. 

Fig.~\ref{fig:contrast}a exhibits an equal partition of the left- and right-localization under $\JL=\JR$, where non-Abelian gauge fields are the only origin of non-Hermiticity. $\eta$ changes sign across the 45- and 135-degree lines defined by $\sin^2(\thL) = \sin^2(\thR)$, exactly the EP phase transition condition. Notably, when $\JL=\JR$, the PBC spectrum collapses into an arc that overlaps with the OBC spectrum. Consequently, the winding number of all OBC energy points is zero, and all modes are extended. 
When $\thL = \left\{0,\pi\right\}$ or $\thR = \left\{0,\pi\right\}$, Eq.~\eqref{eq:HN_nab_E} reduces to the conventional Hatano-Nelson energy band under $\JL=\JR$, whose PBC spectrum also collapses into an arc and all OBC modes are extended; however, there is no phase transition there.

In Fig.~\ref{fig:contrast}b and c, as the imbalance between $\JL$ and $\JR$ appears and increases, localization tunability of the non-Abelian gauge fields becomes suppressed, as shown by the reduced red-colored area.
Crucially, localization tuning is most effective (indicated by color variations in Fig.~\ref{fig:contrast}) near the EP phase transition [Eq.~\eqref{eq:EP} and solid black lines in Fig.~\ref{fig:contrast}].

Asymptotic analysis (Sec.~S4) reveals that the appearance of such tunability stems from the competition between effective nearest-neighbor and high-order [$\left(4N-1\right)$-neighbor where $N$ is an integer] coupling. 
Without loss of generality, we assume $\JL>\JR$ and obtain asymptotic expressions (details in Sec.~S4) of the eigen-energy $E_{\pm}(k)\simeq$
\begin{subequations}
\begin{empheq}[left=\empheqlbrace]{align}
         & \JL\mathrm{e}^{\pm\iu\thL}\mathrm{e}^{\iu k} + \JR\cos{\thR}\mathrm{e}^{-\iu k},  \text{ if } \abs{\JL\sin\thL}\gg\abs{\JR\sin\thR}, 
        \label{eq:asymp_1}\\
         &
         \begin{aligned}
            A(k) \pm \iu \JL\sin{\thL}\mathrm{e}^{\iu k}\sum_{n=0}^{\infty}\mathrm{C}_n^{1/2}\left[\left(\frac{\JR\sin{\thR}}{\JL\sin{\thL}}\right)^2\mathrm{e}^{-\iu4k}\right]^n  \\ 
            \text{ if } \abs{\JL\sin\thL}\gtrapprox\abs{\JR\sin\thR}, 
         \end{aligned}
        \label{eq:asymp_2}\\
       & \begin{aligned}
          A(k) \pm \iu \JR\sin{\thR}\mathrm{e}^{-\iu k} \sum_{n=0}^{\infty} \mathrm{C}_n^{1/2}\left[\left(\frac{\JL\sin{\thL}}{\JR\sin{\thR}}\right)^2\mathrm{e}^{\iu4k}\right]^n \\
          \text{ if } \abs{\JL\sin\thL}\lessapprox\abs{\JR\sin\thR}, 
         \end{aligned}
        \label{eq:asymp_3}\\
         & \JL\cos{\thL}\mathrm{e}^{\iu k} + \JR\mathrm{e}^{\pm\iu\thR}\mathrm{e}^{-\iu k},  \text{ if } \abs{\JL\sin\thL}\ll\abs{\JR\sin\thR},
        \label{eq:asymp_4}
\end{empheq}
\end{subequations}
%
where $\mathrm{C}$ denotes combination.
Away from the EP phase transition [Eqs.~\eqref{eq:asymp_1} and ~\eqref{eq:asymp_4}], the system behaves like a conventional Hatano-Nelson system with modified hopping amplitudes. 
Specifically, a complete left-localization is guaranteed for Eq.~\eqref{eq:asymp_1} because $|\JL|>|\JR\cos\thR|$, while it additionally requires $\abs{\JL\cos\thL}>|\JR|$ for Eq.~\eqref{eq:asymp_4}.  
Near the EP phase transition, higher-order terms become non-negligible, as shown by Eqs.~\eqref{eq:asymp_2} and \eqref{eq:asymp_3} that associate with the two braiding degrees, respectively. 
We plot in Fig.~\ref{fig:contrast}d two gray-shaded regions $|\JL\cos{\thL}| > |\JR|$ $\cup$ $\abs{\JL\sin{\thL}} > \abs{\JR\sin{\thR}}$. Their complementary parameter space is colored in white, near which non-negligible effective high-order hopping occurs. 
The analysis above indicates that the simultaneous presence of NHSE on both ends of the open chain and the localization tunability should appear near the white region, as validated by the numerical calculation in Fig.~\ref{fig:contrast}b.

In conclusion, we introduce and study non-Abelian gauge fields as an origin of non-Hermiticity. In a generalized non-Abelian Hatano--Nelson system under nearest-neighbor coupling only, Hopf-link bulk braiding degrees emerge, whose phase transition accompanies the appearance of EPs. Near the EP phase transition, non-Abelian gauge fields also enable NHSE at both ends of an open chain, and the localization population can be effectively tuned. 

Finally, we discuss the potential experimental platforms to realize non-Abelian gauge fields in non-Hermitian systems. Promising candidates include time-modulated ring resonators (\ie the synthetic frequency dimension)~\cite{dutt2019experimental} and time-multiplexed photonic mesh lattices (\ie the synthetic time dimension)~\cite{regensburger2011photon}. In both cases, we can leverage polarization as the pseudospin degree of freedom. The hopping strength and phases could be engineered by amplitude and phase modulations for each pseudospin component individually to create the desired non-Abelian gauge fields. The experimental observables could be the population of localized modes and their resulting contrast $\eta$ [\eg Eq.~\eqref{eq:contrast}], which can be spectroscopically measured from the intensity of the frequency components or temporally resolved from the pulse intensity at the arrival time that corresponds to the boundaries.

The authors acknowledge the support from the start-up fund of The University of Hong Kong and the National Natural Science Foundation of China Excellent Young Scientists Fund (HKU 12222417).

\providecommand{\noopsort}[1]{}\providecommand{\singleletter}[1]{#1}%


\begin{thebibliography}{74}%
\makeatletter
\providecommand \@ifxundefined [1]{%
 \@ifx{#1\undefined}
}%
\providecommand \@ifnum [1]{%
 \ifnum #1\expandafter \@firstoftwo
 \else \expandafter \@secondoftwo
 \fi
}%
\providecommand \@ifx [1]{%
 \ifx #1\expandafter \@firstoftwo
 \else \expandafter \@secondoftwo
 \fi
}%
\providecommand \natexlab [1]{#1}%
\providecommand \enquote  [1]{``#1''}%
\providecommand \bibnamefont  [1]{#1}%
\providecommand \bibfnamefont [1]{#1}%
\providecommand \citenamefont [1]{#1}%
\providecommand \href@noop [0]{\@secondoftwo}%
\providecommand \href [0]{\begingroup \@sanitize@url \@href}%
\providecommand \@href[1]{\@@startlink{#1}\@@href}%
\providecommand \@@href[1]{\endgroup#1\@@endlink}%
\providecommand \@sanitize@url [0]{\catcode `\\12\catcode `\$12\catcode
  `\&12\catcode `\#12\catcode `\^12\catcode `\_12\catcode `\%12\relax}%
\providecommand \@@startlink[1]{}%
\providecommand \@@endlink[0]{}%
\providecommand \url  [0]{\begingroup\@sanitize@url \@url }%
\providecommand \@url [1]{\endgroup\@href {#1}{\urlprefix }}%
\providecommand \urlprefix  [0]{URL }%
\providecommand \Eprint [0]{\href }%
\providecommand \doibase [0]{https://doi.org/}%
\providecommand \selectlanguage [0]{\@gobble}%
\providecommand \bibinfo  [0]{\@secondoftwo}%
\providecommand \bibfield  [0]{\@secondoftwo}%
\providecommand \translation [1]{[#1]}%
\providecommand \BibitemOpen [0]{}%
\providecommand \bibitemStop [0]{}%
\providecommand \bibitemNoStop [0]{.\EOS\space}%
\providecommand \EOS [0]{\spacefactor3000\relax}%
\providecommand \BibitemShut  [1]{\csname bibitem#1\endcsname}%
\let\auto@bib@innerbib\@empty
\bibitem [{\citenamefont {Musslimani}\ \emph {et~al.}(2008)\citenamefont
  {Musslimani}, \citenamefont {Makris}, \citenamefont {El-Ganainy},\ and\
  \citenamefont {Christodoulides}}]{musslimani2008optical}%
  \BibitemOpen
  \bibfield  {author} {\bibinfo {author} {\bibfnamefont {Z.}~\bibnamefont
  {Musslimani}}, \bibinfo {author} {\bibfnamefont {K.~G.}\ \bibnamefont
  {Makris}}, \bibinfo {author} {\bibfnamefont {R.}~\bibnamefont {El-Ganainy}},\
  and\ \bibinfo {author} {\bibfnamefont {D.~N.}\ \bibnamefont
  {Christodoulides}},\ }\href {https://doi.org/10.1103/PhysRevLett.100.030402}
  {\bibfield  {journal} {\bibinfo  {journal} {Physical Review Letters}\
  }\textbf {\bibinfo {volume} {100}},\ \bibinfo {pages} {030402} (\bibinfo
  {year} {2008})}\BibitemShut {NoStop}%
\bibitem [{\citenamefont {Makris}\ \emph {et~al.}(2008)\citenamefont {Makris},
  \citenamefont {El-Ganainy}, \citenamefont {Christodoulides},\ and\
  \citenamefont {Musslimani}}]{makris2008beam}%
  \BibitemOpen
  \bibfield  {author} {\bibinfo {author} {\bibfnamefont {K.~G.}\ \bibnamefont
  {Makris}}, \bibinfo {author} {\bibfnamefont {R.}~\bibnamefont {El-Ganainy}},
  \bibinfo {author} {\bibfnamefont {D.}~\bibnamefont {Christodoulides}},\ and\
  \bibinfo {author} {\bibfnamefont {Z.~H.}\ \bibnamefont {Musslimani}},\ }\href
  {https://doi.org/10.1103/PhysRevLett.100.103904} {\bibfield  {journal}
  {\bibinfo  {journal} {Physical Review Letters}\ }\textbf {\bibinfo {volume}
  {100}},\ \bibinfo {pages} {103904} (\bibinfo {year} {2008})}\BibitemShut
  {NoStop}%
\bibitem [{\citenamefont {Regensburger}\ \emph {et~al.}(2012)\citenamefont
  {Regensburger}, \citenamefont {Bersch}, \citenamefont {Miri}, \citenamefont
  {Onishchukov}, \citenamefont {Christodoulides},\ and\ \citenamefont
  {Peschel}}]{regensburger2012parity}%
  \BibitemOpen
  \bibfield  {author} {\bibinfo {author} {\bibfnamefont {A.}~\bibnamefont
  {Regensburger}}, \bibinfo {author} {\bibfnamefont {C.}~\bibnamefont
  {Bersch}}, \bibinfo {author} {\bibfnamefont {M.-A.}\ \bibnamefont {Miri}},
  \bibinfo {author} {\bibfnamefont {G.}~\bibnamefont {Onishchukov}}, \bibinfo
  {author} {\bibfnamefont {D.~N.}\ \bibnamefont {Christodoulides}},\ and\
  \bibinfo {author} {\bibfnamefont {U.}~\bibnamefont {Peschel}},\ }\href
  {https://doi.org/10.1038/nature11298} {\bibfield  {journal} {\bibinfo
  {journal} {Nature}\ }\textbf {\bibinfo {volume} {488}},\ \bibinfo {pages}
  {167} (\bibinfo {year} {2012})}\BibitemShut {NoStop}%
\bibitem [{\citenamefont {Feng}\ \emph {et~al.}(2013)\citenamefont {Feng},
  \citenamefont {Xu}, \citenamefont {Fegadolli}, \citenamefont {Lu},
  \citenamefont {Oliveira}, \citenamefont {Almeida}, \citenamefont {Chen},\
  and\ \citenamefont {Scherer}}]{feng2013experimental}%
  \BibitemOpen
  \bibfield  {author} {\bibinfo {author} {\bibfnamefont {L.}~\bibnamefont
  {Feng}}, \bibinfo {author} {\bibfnamefont {Y.-L.}\ \bibnamefont {Xu}},
  \bibinfo {author} {\bibfnamefont {W.~S.}\ \bibnamefont {Fegadolli}}, \bibinfo
  {author} {\bibfnamefont {M.-H.}\ \bibnamefont {Lu}}, \bibinfo {author}
  {\bibfnamefont {J.~E.}\ \bibnamefont {Oliveira}}, \bibinfo {author}
  {\bibfnamefont {V.~R.}\ \bibnamefont {Almeida}}, \bibinfo {author}
  {\bibfnamefont {Y.-F.}\ \bibnamefont {Chen}},\ and\ \bibinfo {author}
  {\bibfnamefont {A.}~\bibnamefont {Scherer}},\ }\href
  {https://doi.org/10.1038/nmat3495} {\bibfield  {journal} {\bibinfo  {journal}
  {Nature materials}\ }\textbf {\bibinfo {volume} {12}},\ \bibinfo {pages}
  {108} (\bibinfo {year} {2013})}\BibitemShut {NoStop}%
\bibitem [{\citenamefont {Gao}\ \emph {et~al.}(2015)\citenamefont {Gao},
  \citenamefont {Estrecho}, \citenamefont {Bliokh}, \citenamefont {Liew},
  \citenamefont {Fraser}, \citenamefont {Brodbeck}, \citenamefont {Kamp},
  \citenamefont {Schneider}, \citenamefont {H{\"o}fling}, \citenamefont
  {Yamamoto} \emph {et~al.}}]{gao2015observation}%
  \BibitemOpen
  \bibfield  {author} {\bibinfo {author} {\bibfnamefont {T.}~\bibnamefont
  {Gao}}, \bibinfo {author} {\bibfnamefont {E.}~\bibnamefont {Estrecho}},
  \bibinfo {author} {\bibfnamefont {K.}~\bibnamefont {Bliokh}}, \bibinfo
  {author} {\bibfnamefont {T.}~\bibnamefont {Liew}}, \bibinfo {author}
  {\bibfnamefont {M.}~\bibnamefont {Fraser}}, \bibinfo {author} {\bibfnamefont
  {S.}~\bibnamefont {Brodbeck}}, \bibinfo {author} {\bibfnamefont
  {M.}~\bibnamefont {Kamp}}, \bibinfo {author} {\bibfnamefont {C.}~\bibnamefont
  {Schneider}}, \bibinfo {author} {\bibfnamefont {S.}~\bibnamefont
  {H{\"o}fling}}, \bibinfo {author} {\bibfnamefont {Y.}~\bibnamefont
  {Yamamoto}}, \emph {et~al.},\ }\href {https://doi.org/10.1038/nature15522}
  {\bibfield  {journal} {\bibinfo  {journal} {Nature}\ }\textbf {\bibinfo
  {volume} {526}},\ \bibinfo {pages} {554} (\bibinfo {year}
  {2015})}\BibitemShut {NoStop}%
\bibitem [{\citenamefont {Doppler}\ \emph {et~al.}(2016)\citenamefont
  {Doppler}, \citenamefont {Mailybaev}, \citenamefont {B{\"o}hm}, \citenamefont
  {Kuhl}, \citenamefont {Girschik}, \citenamefont {Libisch}, \citenamefont
  {Milburn}, \citenamefont {Rabl}, \citenamefont {Moiseyev},\ and\
  \citenamefont {Rotter}}]{doppler2016dynamically}%
  \BibitemOpen
  \bibfield  {author} {\bibinfo {author} {\bibfnamefont {J.}~\bibnamefont
  {Doppler}}, \bibinfo {author} {\bibfnamefont {A.~A.}\ \bibnamefont
  {Mailybaev}}, \bibinfo {author} {\bibfnamefont {J.}~\bibnamefont {B{\"o}hm}},
  \bibinfo {author} {\bibfnamefont {U.}~\bibnamefont {Kuhl}}, \bibinfo {author}
  {\bibfnamefont {A.}~\bibnamefont {Girschik}}, \bibinfo {author}
  {\bibfnamefont {F.}~\bibnamefont {Libisch}}, \bibinfo {author} {\bibfnamefont
  {T.~J.}\ \bibnamefont {Milburn}}, \bibinfo {author} {\bibfnamefont
  {P.}~\bibnamefont {Rabl}}, \bibinfo {author} {\bibfnamefont {N.}~\bibnamefont
  {Moiseyev}},\ and\ \bibinfo {author} {\bibfnamefont {S.}~\bibnamefont
  {Rotter}},\ }\href {https://doi.org/10.1038/nature18605} {\bibfield
  {journal} {\bibinfo  {journal} {Nature}\ }\textbf {\bibinfo {volume} {537}},\
  \bibinfo {pages} {76} (\bibinfo {year} {2016})}\BibitemShut {NoStop}%
\bibitem [{\citenamefont {Yao}\ and\ \citenamefont {Wang}(2018)}]{yao2018edge}%
  \BibitemOpen
  \bibfield  {author} {\bibinfo {author} {\bibfnamefont {S.}~\bibnamefont
  {Yao}}\ and\ \bibinfo {author} {\bibfnamefont {Z.}~\bibnamefont {Wang}},\
  }\href {https://doi.org/10.1103/PhysRevLett.121.086803} {\bibfield  {journal}
  {\bibinfo  {journal} {Physical review letters}\ }\textbf {\bibinfo {volume}
  {121}},\ \bibinfo {pages} {086803} (\bibinfo {year} {2018})}\BibitemShut
  {NoStop}%
\bibitem [{\citenamefont {Song}\ \emph {et~al.}(2019)\citenamefont {Song},
  \citenamefont {Yao},\ and\ \citenamefont {Wang}}]{song2019non}%
  \BibitemOpen
  \bibfield  {author} {\bibinfo {author} {\bibfnamefont {F.}~\bibnamefont
  {Song}}, \bibinfo {author} {\bibfnamefont {S.}~\bibnamefont {Yao}},\ and\
  \bibinfo {author} {\bibfnamefont {Z.}~\bibnamefont {Wang}},\ }\href
  {https://doi.org/10.1103/PhysRevLett.123.246801} {\bibfield  {journal}
  {\bibinfo  {journal} {Physical review letters}\ }\textbf {\bibinfo {volume}
  {123}},\ \bibinfo {pages} {246801} (\bibinfo {year} {2019})}\BibitemShut
  {NoStop}%
\bibitem [{\citenamefont {Kunst}\ \emph {et~al.}(2018)\citenamefont {Kunst},
  \citenamefont {Edvardsson}, \citenamefont {Budich},\ and\ \citenamefont
  {Bergholtz}}]{kunst2018biorthogonal}%
  \BibitemOpen
  \bibfield  {author} {\bibinfo {author} {\bibfnamefont {F.~K.}\ \bibnamefont
  {Kunst}}, \bibinfo {author} {\bibfnamefont {E.}~\bibnamefont {Edvardsson}},
  \bibinfo {author} {\bibfnamefont {J.~C.}\ \bibnamefont {Budich}},\ and\
  \bibinfo {author} {\bibfnamefont {E.~J.}\ \bibnamefont {Bergholtz}},\ }\href
  {https://doi.org/10.1103/PhysRevLett.121.026808} {\bibfield  {journal}
  {\bibinfo  {journal} {Physical review letters}\ }\textbf {\bibinfo {volume}
  {121}},\ \bibinfo {pages} {026808} (\bibinfo {year} {2018})}\BibitemShut
  {NoStop}%
\bibitem [{\citenamefont {Xiong}(2018)}]{xiong2018does}%
  \BibitemOpen
  \bibfield  {author} {\bibinfo {author} {\bibfnamefont {Y.}~\bibnamefont
  {Xiong}},\ }\href {https://doi.org/10.1088/2399-6528/aab64a} {\bibfield
  {journal} {\bibinfo  {journal} {Journal of Physics Communications}\ }\textbf
  {\bibinfo {volume} {2}},\ \bibinfo {pages} {035043} (\bibinfo {year}
  {2018})}\BibitemShut {NoStop}%
\bibitem [{\citenamefont {Bergholtz}\ \emph {et~al.}(2021)\citenamefont
  {Bergholtz}, \citenamefont {Budich},\ and\ \citenamefont
  {Kunst}}]{bergholtz2021exceptional}%
  \BibitemOpen
  \bibfield  {author} {\bibinfo {author} {\bibfnamefont {E.~J.}\ \bibnamefont
  {Bergholtz}}, \bibinfo {author} {\bibfnamefont {J.~C.}\ \bibnamefont
  {Budich}},\ and\ \bibinfo {author} {\bibfnamefont {F.~K.}\ \bibnamefont
  {Kunst}},\ }\href {https://doi.org/10.1103/RevModPhys.93.015005} {\bibfield
  {journal} {\bibinfo  {journal} {Reviews of Modern Physics}\ }\textbf
  {\bibinfo {volume} {93}},\ \bibinfo {pages} {015005} (\bibinfo {year}
  {2021})}\BibitemShut {NoStop}%
\bibitem [{\citenamefont {Gong}\ \emph {et~al.}(2018)\citenamefont {Gong},
  \citenamefont {Ashida}, \citenamefont {Kawabata}, \citenamefont {Takasan},
  \citenamefont {Higashikawa},\ and\ \citenamefont
  {Ueda}}]{gong2018topological}%
  \BibitemOpen
  \bibfield  {author} {\bibinfo {author} {\bibfnamefont {Z.}~\bibnamefont
  {Gong}}, \bibinfo {author} {\bibfnamefont {Y.}~\bibnamefont {Ashida}},
  \bibinfo {author} {\bibfnamefont {K.}~\bibnamefont {Kawabata}}, \bibinfo
  {author} {\bibfnamefont {K.}~\bibnamefont {Takasan}}, \bibinfo {author}
  {\bibfnamefont {S.}~\bibnamefont {Higashikawa}},\ and\ \bibinfo {author}
  {\bibfnamefont {M.}~\bibnamefont {Ueda}},\ }\href
  {https://doi.org/10.1103/PhysRevX.8.031079} {\bibfield  {journal} {\bibinfo
  {journal} {Physical Review X}\ }\textbf {\bibinfo {volume} {8}},\ \bibinfo
  {pages} {031079} (\bibinfo {year} {2018})}\BibitemShut {NoStop}%
\bibitem [{\citenamefont {Kawabata}\ \emph {et~al.}(2019)\citenamefont
  {Kawabata}, \citenamefont {Shiozaki}, \citenamefont {Ueda},\ and\
  \citenamefont {Sato}}]{kawabata2019symmetry}%
  \BibitemOpen
  \bibfield  {author} {\bibinfo {author} {\bibfnamefont {K.}~\bibnamefont
  {Kawabata}}, \bibinfo {author} {\bibfnamefont {K.}~\bibnamefont {Shiozaki}},
  \bibinfo {author} {\bibfnamefont {M.}~\bibnamefont {Ueda}},\ and\ \bibinfo
  {author} {\bibfnamefont {M.}~\bibnamefont {Sato}},\ }\href
  {https://doi.org/10.1103/PhysRevX.9.041015} {\bibfield  {journal} {\bibinfo
  {journal} {Physical Review X}\ }\textbf {\bibinfo {volume} {9}},\ \bibinfo
  {pages} {041015} (\bibinfo {year} {2019})}\BibitemShut {NoStop}%
\bibitem [{\citenamefont {Yokomizo}\ and\ \citenamefont
  {Murakami}(2019)}]{yokomizo2019non}%
  \BibitemOpen
  \bibfield  {author} {\bibinfo {author} {\bibfnamefont {K.}~\bibnamefont
  {Yokomizo}}\ and\ \bibinfo {author} {\bibfnamefont {S.}~\bibnamefont
  {Murakami}},\ }\href {https://doi.org/10.1103/PhysRevLett.123.066404}
  {\bibfield  {journal} {\bibinfo  {journal} {Physical review letters}\
  }\textbf {\bibinfo {volume} {123}},\ \bibinfo {pages} {066404} (\bibinfo
  {year} {2019})}\BibitemShut {NoStop}%
\bibitem [{\citenamefont {Zhang}\ \emph {et~al.}(2020)\citenamefont {Zhang},
  \citenamefont {Yang},\ and\ \citenamefont {Fang}}]{zhang2020correspondence}%
  \BibitemOpen
  \bibfield  {author} {\bibinfo {author} {\bibfnamefont {K.}~\bibnamefont
  {Zhang}}, \bibinfo {author} {\bibfnamefont {Z.}~\bibnamefont {Yang}},\ and\
  \bibinfo {author} {\bibfnamefont {C.}~\bibnamefont {Fang}},\ }\href
  {https://doi.org/10.1103/PhysRevLett.125.126402} {\bibfield  {journal}
  {\bibinfo  {journal} {Physical Review Letters}\ }\textbf {\bibinfo {volume}
  {125}},\ \bibinfo {pages} {126402} (\bibinfo {year} {2020})}\BibitemShut
  {NoStop}%
\bibitem [{\citenamefont {Okuma}\ \emph {et~al.}(2020)\citenamefont {Okuma},
  \citenamefont {Kawabata}, \citenamefont {Shiozaki},\ and\ \citenamefont
  {Sato}}]{okuma2020topological}%
  \BibitemOpen
  \bibfield  {author} {\bibinfo {author} {\bibfnamefont {N.}~\bibnamefont
  {Okuma}}, \bibinfo {author} {\bibfnamefont {K.}~\bibnamefont {Kawabata}},
  \bibinfo {author} {\bibfnamefont {K.}~\bibnamefont {Shiozaki}},\ and\
  \bibinfo {author} {\bibfnamefont {M.}~\bibnamefont {Sato}},\ }\href
  {https://doi.org/10.1103/PhysRevLett.124.086801} {\bibfield  {journal}
  {\bibinfo  {journal} {Physical review letters}\ }\textbf {\bibinfo {volume}
  {124}},\ \bibinfo {pages} {086801} (\bibinfo {year} {2020})}\BibitemShut
  {NoStop}%
\bibitem [{\citenamefont {Yang}\ \emph
  {et~al.}(2020{\natexlab{a}})\citenamefont {Yang}, \citenamefont {Zhang},
  \citenamefont {Fang},\ and\ \citenamefont {Hu}}]{yang2020non}%
  \BibitemOpen
  \bibfield  {author} {\bibinfo {author} {\bibfnamefont {Z.}~\bibnamefont
  {Yang}}, \bibinfo {author} {\bibfnamefont {K.}~\bibnamefont {Zhang}},
  \bibinfo {author} {\bibfnamefont {C.}~\bibnamefont {Fang}},\ and\ \bibinfo
  {author} {\bibfnamefont {J.}~\bibnamefont {Hu}},\ }\href
  {https://doi.org/10.1103/PhysRevLett.125.226402} {\bibfield  {journal}
  {\bibinfo  {journal} {Physical Review Letters}\ }\textbf {\bibinfo {volume}
  {125}},\ \bibinfo {pages} {226402} (\bibinfo {year}
  {2020}{\natexlab{a}})}\BibitemShut {NoStop}%
\bibitem [{\citenamefont {Wang}\ \emph
  {et~al.}(2021{\natexlab{a}})\citenamefont {Wang}, \citenamefont {Dutt},
  \citenamefont {Wojcik},\ and\ \citenamefont {Fan}}]{wang2021topological}%
  \BibitemOpen
  \bibfield  {author} {\bibinfo {author} {\bibfnamefont {K.}~\bibnamefont
  {Wang}}, \bibinfo {author} {\bibfnamefont {A.}~\bibnamefont {Dutt}}, \bibinfo
  {author} {\bibfnamefont {C.~C.}\ \bibnamefont {Wojcik}},\ and\ \bibinfo
  {author} {\bibfnamefont {S.}~\bibnamefont {Fan}},\ }\href
  {https://doi.org/10.1038/s41586-021-03848-x} {\bibfield  {journal} {\bibinfo
  {journal} {Nature}\ }\textbf {\bibinfo {volume} {598}},\ \bibinfo {pages}
  {59} (\bibinfo {year} {2021}{\natexlab{a}})}\BibitemShut {NoStop}%
\bibitem [{\citenamefont {Hu}\ and\ \citenamefont {Zhao}(2021)}]{hu2021knots}%
  \BibitemOpen
  \bibfield  {author} {\bibinfo {author} {\bibfnamefont {H.}~\bibnamefont
  {Hu}}\ and\ \bibinfo {author} {\bibfnamefont {E.}~\bibnamefont {Zhao}},\
  }\href {https://doi.org/10.1103/PhysRevLett.126.010401} {\bibfield  {journal}
  {\bibinfo  {journal} {Physical Review Letters}\ }\textbf {\bibinfo {volume}
  {126}},\ \bibinfo {pages} {010401} (\bibinfo {year} {2021})}\BibitemShut
  {NoStop}%
\bibitem [{\citenamefont {Borgnia}\ \emph {et~al.}(2020)\citenamefont
  {Borgnia}, \citenamefont {Kruchkov},\ and\ \citenamefont
  {Slager}}]{borgnia2020non}%
  \BibitemOpen
  \bibfield  {author} {\bibinfo {author} {\bibfnamefont {D.~S.}\ \bibnamefont
  {Borgnia}}, \bibinfo {author} {\bibfnamefont {A.~J.}\ \bibnamefont
  {Kruchkov}},\ and\ \bibinfo {author} {\bibfnamefont {R.-J.}\ \bibnamefont
  {Slager}},\ }\href {https://doi.org/10.1103/PhysRevLett.124.056802}
  {\bibfield  {journal} {\bibinfo  {journal} {Physical review letters}\
  }\textbf {\bibinfo {volume} {124}},\ \bibinfo {pages} {056802} (\bibinfo
  {year} {2020})}\BibitemShut {NoStop}%
\bibitem [{\citenamefont {Lee}(2016)}]{lee2016anomalous}%
  \BibitemOpen
  \bibfield  {author} {\bibinfo {author} {\bibfnamefont {T.~E.}\ \bibnamefont
  {Lee}},\ }\href {https://doi.org/10.1103/PhysRevLett.116.133903} {\bibfield
  {journal} {\bibinfo  {journal} {Physical review letters}\ }\textbf {\bibinfo
  {volume} {116}},\ \bibinfo {pages} {133903} (\bibinfo {year}
  {2016})}\BibitemShut {NoStop}%
\bibitem [{\citenamefont {Weidemann}\ \emph {et~al.}(2020)\citenamefont
  {Weidemann}, \citenamefont {Kremer}, \citenamefont {Helbig}, \citenamefont
  {Hofmann}, \citenamefont {Stegmaier}, \citenamefont {Greiter}, \citenamefont
  {Thomale},\ and\ \citenamefont {Szameit}}]{weidemann2020topological}%
  \BibitemOpen
  \bibfield  {author} {\bibinfo {author} {\bibfnamefont {S.}~\bibnamefont
  {Weidemann}}, \bibinfo {author} {\bibfnamefont {M.}~\bibnamefont {Kremer}},
  \bibinfo {author} {\bibfnamefont {T.}~\bibnamefont {Helbig}}, \bibinfo
  {author} {\bibfnamefont {T.}~\bibnamefont {Hofmann}}, \bibinfo {author}
  {\bibfnamefont {A.}~\bibnamefont {Stegmaier}}, \bibinfo {author}
  {\bibfnamefont {M.}~\bibnamefont {Greiter}}, \bibinfo {author} {\bibfnamefont
  {R.}~\bibnamefont {Thomale}},\ and\ \bibinfo {author} {\bibfnamefont
  {A.}~\bibnamefont {Szameit}},\ }\href
  {https://doi.org/10.1126/science.aaz8727} {\bibfield  {journal} {\bibinfo
  {journal} {Science}\ }\textbf {\bibinfo {volume} {368}},\ \bibinfo {pages}
  {311} (\bibinfo {year} {2020})}\BibitemShut {NoStop}%
\bibitem [{\citenamefont {Wang}\ \emph
  {et~al.}(2021{\natexlab{b}})\citenamefont {Wang}, \citenamefont {Dutt},
  \citenamefont {Yang}, \citenamefont {Wojcik}, \citenamefont
  {Vu{\v{c}}kovi{\'c}},\ and\ \citenamefont {Fan}}]{wang2021generating}%
  \BibitemOpen
  \bibfield  {author} {\bibinfo {author} {\bibfnamefont {K.}~\bibnamefont
  {Wang}}, \bibinfo {author} {\bibfnamefont {A.}~\bibnamefont {Dutt}}, \bibinfo
  {author} {\bibfnamefont {K.~Y.}\ \bibnamefont {Yang}}, \bibinfo {author}
  {\bibfnamefont {C.~C.}\ \bibnamefont {Wojcik}}, \bibinfo {author}
  {\bibfnamefont {J.}~\bibnamefont {Vu{\v{c}}kovi{\'c}}},\ and\ \bibinfo
  {author} {\bibfnamefont {S.}~\bibnamefont {Fan}},\ }\href
  {https://doi.org/10.1126/science.abf6568} {\bibfield  {journal} {\bibinfo
  {journal} {Science}\ }\textbf {\bibinfo {volume} {371}},\ \bibinfo {pages}
  {1240} (\bibinfo {year} {2021}{\natexlab{b}})}\BibitemShut {NoStop}%
\bibitem [{\citenamefont {Xiao}\ \emph {et~al.}(2020)\citenamefont {Xiao},
  \citenamefont {Deng}, \citenamefont {Wang}, \citenamefont {Zhu},
  \citenamefont {Wang}, \citenamefont {Yi},\ and\ \citenamefont
  {Xue}}]{xiao2020non}%
  \BibitemOpen
  \bibfield  {author} {\bibinfo {author} {\bibfnamefont {L.}~\bibnamefont
  {Xiao}}, \bibinfo {author} {\bibfnamefont {T.}~\bibnamefont {Deng}}, \bibinfo
  {author} {\bibfnamefont {K.}~\bibnamefont {Wang}}, \bibinfo {author}
  {\bibfnamefont {G.}~\bibnamefont {Zhu}}, \bibinfo {author} {\bibfnamefont
  {Z.}~\bibnamefont {Wang}}, \bibinfo {author} {\bibfnamefont {W.}~\bibnamefont
  {Yi}},\ and\ \bibinfo {author} {\bibfnamefont {P.}~\bibnamefont {Xue}},\
  }\href {https://doi.org/10.1038/s41567-020-0836-6} {\bibfield  {journal}
  {\bibinfo  {journal} {Nature Physics}\ }\textbf {\bibinfo {volume} {16}},\
  \bibinfo {pages} {761} (\bibinfo {year} {2020})}\BibitemShut {NoStop}%
\bibitem [{\citenamefont {Liu}\ \emph {et~al.}(2022)\citenamefont {Liu},
  \citenamefont {Wei}, \citenamefont {Hemmatyar}, \citenamefont {Pyrialakos},
  \citenamefont {Jung}, \citenamefont {Christodoulides},\ and\ \citenamefont
  {Khajavikhan}}]{liu2022complex}%
  \BibitemOpen
  \bibfield  {author} {\bibinfo {author} {\bibfnamefont {Y.~G.}\ \bibnamefont
  {Liu}}, \bibinfo {author} {\bibfnamefont {Y.}~\bibnamefont {Wei}}, \bibinfo
  {author} {\bibfnamefont {O.}~\bibnamefont {Hemmatyar}}, \bibinfo {author}
  {\bibfnamefont {G.~G.}\ \bibnamefont {Pyrialakos}}, \bibinfo {author}
  {\bibfnamefont {P.~S.}\ \bibnamefont {Jung}}, \bibinfo {author}
  {\bibfnamefont {D.~N.}\ \bibnamefont {Christodoulides}},\ and\ \bibinfo
  {author} {\bibfnamefont {M.}~\bibnamefont {Khajavikhan}},\ }\href
  {https://doi.org/10.1038/s41377-022-01030-0} {\bibfield  {journal} {\bibinfo
  {journal} {Light: Science \& Applications}\ }\textbf {\bibinfo {volume}
  {11}},\ \bibinfo {pages} {336} (\bibinfo {year} {2022})}\BibitemShut
  {NoStop}%
\bibitem [{\citenamefont {Zhang}\ \emph
  {et~al.}(2021{\natexlab{a}})\citenamefont {Zhang}, \citenamefont {Tian},
  \citenamefont {Jiang}, \citenamefont {Lu},\ and\ \citenamefont
  {Chen}}]{zhang2021observation}%
  \BibitemOpen
  \bibfield  {author} {\bibinfo {author} {\bibfnamefont {X.}~\bibnamefont
  {Zhang}}, \bibinfo {author} {\bibfnamefont {Y.}~\bibnamefont {Tian}},
  \bibinfo {author} {\bibfnamefont {J.-H.}\ \bibnamefont {Jiang}}, \bibinfo
  {author} {\bibfnamefont {M.-H.}\ \bibnamefont {Lu}},\ and\ \bibinfo {author}
  {\bibfnamefont {Y.-F.}\ \bibnamefont {Chen}},\ }\href
  {https://doi.org/10.1038/s41467-021-25716-y} {\bibfield  {journal} {\bibinfo
  {journal} {Nature communications}\ }\textbf {\bibinfo {volume} {12}},\
  \bibinfo {pages} {5377} (\bibinfo {year} {2021}{\natexlab{a}})}\BibitemShut
  {NoStop}%
\bibitem [{\citenamefont {Zhang}\ \emph
  {et~al.}(2021{\natexlab{b}})\citenamefont {Zhang}, \citenamefont {Yang},
  \citenamefont {Ge}, \citenamefont {Guan}, \citenamefont {Chen}, \citenamefont
  {Yan}, \citenamefont {Chen}, \citenamefont {Xi}, \citenamefont {Li},
  \citenamefont {Jia} \emph {et~al.}}]{zhang2021acoustic}%
  \BibitemOpen
  \bibfield  {author} {\bibinfo {author} {\bibfnamefont {L.}~\bibnamefont
  {Zhang}}, \bibinfo {author} {\bibfnamefont {Y.}~\bibnamefont {Yang}},
  \bibinfo {author} {\bibfnamefont {Y.}~\bibnamefont {Ge}}, \bibinfo {author}
  {\bibfnamefont {Y.-J.}\ \bibnamefont {Guan}}, \bibinfo {author}
  {\bibfnamefont {Q.}~\bibnamefont {Chen}}, \bibinfo {author} {\bibfnamefont
  {Q.}~\bibnamefont {Yan}}, \bibinfo {author} {\bibfnamefont {F.}~\bibnamefont
  {Chen}}, \bibinfo {author} {\bibfnamefont {R.}~\bibnamefont {Xi}}, \bibinfo
  {author} {\bibfnamefont {Y.}~\bibnamefont {Li}}, \bibinfo {author}
  {\bibfnamefont {D.}~\bibnamefont {Jia}}, \emph {et~al.},\ }\href
  {https://doi.org/10.1038/s41467-021-26619-8} {\bibfield  {journal} {\bibinfo
  {journal} {Nature communications}\ }\textbf {\bibinfo {volume} {12}},\
  \bibinfo {pages} {6297} (\bibinfo {year} {2021}{\natexlab{b}})}\BibitemShut
  {NoStop}%
\bibitem [{\citenamefont {Gu}\ \emph {et~al.}(2022)\citenamefont {Gu},
  \citenamefont {Gao}, \citenamefont {Xue}, \citenamefont {Li}, \citenamefont
  {Su},\ and\ \citenamefont {Zhu}}]{gu2022transient}%
  \BibitemOpen
  \bibfield  {author} {\bibinfo {author} {\bibfnamefont {Z.}~\bibnamefont
  {Gu}}, \bibinfo {author} {\bibfnamefont {H.}~\bibnamefont {Gao}}, \bibinfo
  {author} {\bibfnamefont {H.}~\bibnamefont {Xue}}, \bibinfo {author}
  {\bibfnamefont {J.}~\bibnamefont {Li}}, \bibinfo {author} {\bibfnamefont
  {Z.}~\bibnamefont {Su}},\ and\ \bibinfo {author} {\bibfnamefont
  {J.}~\bibnamefont {Zhu}},\ }\href
  {https://doi.org/10.1038/s41467-022-35448-2} {\bibfield  {journal} {\bibinfo
  {journal} {Nature Communications}\ }\textbf {\bibinfo {volume} {13}},\
  \bibinfo {pages} {7668} (\bibinfo {year} {2022})}\BibitemShut {NoStop}%
\bibitem [{\citenamefont {Wang}\ \emph {et~al.}(2022)\citenamefont {Wang},
  \citenamefont {Wang},\ and\ \citenamefont {Ma}}]{wang2022non}%
  \BibitemOpen
  \bibfield  {author} {\bibinfo {author} {\bibfnamefont {W.}~\bibnamefont
  {Wang}}, \bibinfo {author} {\bibfnamefont {X.}~\bibnamefont {Wang}},\ and\
  \bibinfo {author} {\bibfnamefont {G.}~\bibnamefont {Ma}},\ }\href
  {https://doi.org/10.1038/s41586-022-04929-1} {\bibfield  {journal} {\bibinfo
  {journal} {Nature}\ }\textbf {\bibinfo {volume} {608}},\ \bibinfo {pages}
  {50} (\bibinfo {year} {2022})}\BibitemShut {NoStop}%
\bibitem [{\citenamefont {Ghatak}\ \emph {et~al.}(2020)\citenamefont {Ghatak},
  \citenamefont {Brandenbourger}, \citenamefont {Van~Wezel},\ and\
  \citenamefont {Coulais}}]{ghatak2020observation}%
  \BibitemOpen
  \bibfield  {author} {\bibinfo {author} {\bibfnamefont {A.}~\bibnamefont
  {Ghatak}}, \bibinfo {author} {\bibfnamefont {M.}~\bibnamefont
  {Brandenbourger}}, \bibinfo {author} {\bibfnamefont {J.}~\bibnamefont
  {Van~Wezel}},\ and\ \bibinfo {author} {\bibfnamefont {C.}~\bibnamefont
  {Coulais}},\ }\href {https://doi.org/10.1073/pnas.2010580117} {\bibfield
  {journal} {\bibinfo  {journal} {Proceedings of the National Academy of
  Sciences}\ }\textbf {\bibinfo {volume} {117}},\ \bibinfo {pages} {29561}
  (\bibinfo {year} {2020})}\BibitemShut {NoStop}%
\bibitem [{\citenamefont {Scheibner}\ \emph {et~al.}(2020)\citenamefont
  {Scheibner}, \citenamefont {Irvine},\ and\ \citenamefont
  {Vitelli}}]{scheibner2020non}%
  \BibitemOpen
  \bibfield  {author} {\bibinfo {author} {\bibfnamefont {C.}~\bibnamefont
  {Scheibner}}, \bibinfo {author} {\bibfnamefont {W.~T.}\ \bibnamefont
  {Irvine}},\ and\ \bibinfo {author} {\bibfnamefont {V.}~\bibnamefont
  {Vitelli}},\ }\href {https://doi.org/10.1103/PhysRevLett.125.118001}
  {\bibfield  {journal} {\bibinfo  {journal} {Physical Review Letters}\
  }\textbf {\bibinfo {volume} {125}},\ \bibinfo {pages} {118001} (\bibinfo
  {year} {2020})}\BibitemShut {NoStop}%
\bibitem [{\citenamefont {Li}\ \emph {et~al.}(2020)\citenamefont {Li},
  \citenamefont {Lee}, \citenamefont {Mu},\ and\ \citenamefont
  {Gong}}]{li2020critical}%
  \BibitemOpen
  \bibfield  {author} {\bibinfo {author} {\bibfnamefont {L.}~\bibnamefont
  {Li}}, \bibinfo {author} {\bibfnamefont {C.~H.}\ \bibnamefont {Lee}},
  \bibinfo {author} {\bibfnamefont {S.}~\bibnamefont {Mu}},\ and\ \bibinfo
  {author} {\bibfnamefont {J.}~\bibnamefont {Gong}},\ }\href
  {https://doi.org/10.1038/s41467-020-18917-4} {\bibfield  {journal} {\bibinfo
  {journal} {Nature communications}\ }\textbf {\bibinfo {volume} {11}},\
  \bibinfo {pages} {5491} (\bibinfo {year} {2020})}\BibitemShut {NoStop}%
\bibitem [{\citenamefont {Helbig}\ \emph {et~al.}(2020)\citenamefont {Helbig},
  \citenamefont {Hofmann}, \citenamefont {Imhof}, \citenamefont {Abdelghany},
  \citenamefont {Kiessling}, \citenamefont {Molenkamp}, \citenamefont {Lee},
  \citenamefont {Szameit}, \citenamefont {Greiter},\ and\ \citenamefont
  {Thomale}}]{helbig2020generalized}%
  \BibitemOpen
  \bibfield  {author} {\bibinfo {author} {\bibfnamefont {T.}~\bibnamefont
  {Helbig}}, \bibinfo {author} {\bibfnamefont {T.}~\bibnamefont {Hofmann}},
  \bibinfo {author} {\bibfnamefont {S.}~\bibnamefont {Imhof}}, \bibinfo
  {author} {\bibfnamefont {M.}~\bibnamefont {Abdelghany}}, \bibinfo {author}
  {\bibfnamefont {T.}~\bibnamefont {Kiessling}}, \bibinfo {author}
  {\bibfnamefont {L.}~\bibnamefont {Molenkamp}}, \bibinfo {author}
  {\bibfnamefont {C.}~\bibnamefont {Lee}}, \bibinfo {author} {\bibfnamefont
  {A.}~\bibnamefont {Szameit}}, \bibinfo {author} {\bibfnamefont
  {M.}~\bibnamefont {Greiter}},\ and\ \bibinfo {author} {\bibfnamefont
  {R.}~\bibnamefont {Thomale}},\ }\href
  {https://doi.org/10.1038/s41567-020-0922-9} {\bibfield  {journal} {\bibinfo
  {journal} {Nature Physics}\ }\textbf {\bibinfo {volume} {16}},\ \bibinfo
  {pages} {747} (\bibinfo {year} {2020})}\BibitemShut {NoStop}%
\bibitem [{\citenamefont {Hofmann}\ \emph {et~al.}(2020)\citenamefont
  {Hofmann}, \citenamefont {Helbig}, \citenamefont {Schindler}, \citenamefont
  {Salgo}, \citenamefont {Brzezi{\'n}ska}, \citenamefont {Greiter},
  \citenamefont {Kiessling}, \citenamefont {Wolf}, \citenamefont {Vollhardt},
  \citenamefont {Kaba{\v{s}}i} \emph {et~al.}}]{hofmann2020reciprocal}%
  \BibitemOpen
  \bibfield  {author} {\bibinfo {author} {\bibfnamefont {T.}~\bibnamefont
  {Hofmann}}, \bibinfo {author} {\bibfnamefont {T.}~\bibnamefont {Helbig}},
  \bibinfo {author} {\bibfnamefont {F.}~\bibnamefont {Schindler}}, \bibinfo
  {author} {\bibfnamefont {N.}~\bibnamefont {Salgo}}, \bibinfo {author}
  {\bibfnamefont {M.}~\bibnamefont {Brzezi{\'n}ska}}, \bibinfo {author}
  {\bibfnamefont {M.}~\bibnamefont {Greiter}}, \bibinfo {author} {\bibfnamefont
  {T.}~\bibnamefont {Kiessling}}, \bibinfo {author} {\bibfnamefont
  {D.}~\bibnamefont {Wolf}}, \bibinfo {author} {\bibfnamefont {A.}~\bibnamefont
  {Vollhardt}}, \bibinfo {author} {\bibfnamefont {A.}~\bibnamefont
  {Kaba{\v{s}}i}}, \emph {et~al.},\ }\href
  {https://doi.org/10.1103/PhysRevResearch.2.023265} {\bibfield  {journal}
  {\bibinfo  {journal} {Physical Review Research}\ }\textbf {\bibinfo {volume}
  {2}},\ \bibinfo {pages} {023265} (\bibinfo {year} {2020})}\BibitemShut
  {NoStop}%
\bibitem [{\citenamefont {Liu}\ \emph {et~al.}(2021)\citenamefont {Liu},
  \citenamefont {Shao}, \citenamefont {Ma}, \citenamefont {Zhang},
  \citenamefont {You}, \citenamefont {Wu}, \citenamefont {Xiang}, \citenamefont
  {Cui},\ and\ \citenamefont {Zhang}}]{liu2021non}%
  \BibitemOpen
  \bibfield  {author} {\bibinfo {author} {\bibfnamefont {S.}~\bibnamefont
  {Liu}}, \bibinfo {author} {\bibfnamefont {R.}~\bibnamefont {Shao}}, \bibinfo
  {author} {\bibfnamefont {S.}~\bibnamefont {Ma}}, \bibinfo {author}
  {\bibfnamefont {L.}~\bibnamefont {Zhang}}, \bibinfo {author} {\bibfnamefont
  {O.}~\bibnamefont {You}}, \bibinfo {author} {\bibfnamefont {H.}~\bibnamefont
  {Wu}}, \bibinfo {author} {\bibfnamefont {Y.~J.}\ \bibnamefont {Xiang}},
  \bibinfo {author} {\bibfnamefont {T.~J.}\ \bibnamefont {Cui}},\ and\ \bibinfo
  {author} {\bibfnamefont {S.}~\bibnamefont {Zhang}},\ }\href
  {https://doi.org/10.34133/2021/5608038} {\bibfield  {journal} {\bibinfo
  {journal} {Research}\ ,\ \bibinfo {pages} {5608038}} (\bibinfo {year}
  {2021})}\BibitemShut {NoStop}%
\bibitem [{\citenamefont {Zou}\ \emph {et~al.}(2021)\citenamefont {Zou},
  \citenamefont {Chen}, \citenamefont {He}, \citenamefont {Bao}, \citenamefont
  {Lee}, \citenamefont {Sun},\ and\ \citenamefont
  {Zhang}}]{zou2021observation}%
  \BibitemOpen
  \bibfield  {author} {\bibinfo {author} {\bibfnamefont {D.}~\bibnamefont
  {Zou}}, \bibinfo {author} {\bibfnamefont {T.}~\bibnamefont {Chen}}, \bibinfo
  {author} {\bibfnamefont {W.}~\bibnamefont {He}}, \bibinfo {author}
  {\bibfnamefont {J.}~\bibnamefont {Bao}}, \bibinfo {author} {\bibfnamefont
  {C.~H.}\ \bibnamefont {Lee}}, \bibinfo {author} {\bibfnamefont
  {H.}~\bibnamefont {Sun}},\ and\ \bibinfo {author} {\bibfnamefont
  {X.}~\bibnamefont {Zhang}},\ }\href
  {https://doi.org/10.1038/s41467-021-26414-5} {\bibfield  {journal} {\bibinfo
  {journal} {Nature Communications}\ }\textbf {\bibinfo {volume} {12}},\
  \bibinfo {pages} {7201} (\bibinfo {year} {2021})}\BibitemShut {NoStop}%
\bibitem [{\citenamefont {Shang}\ \emph {et~al.}(2022)\citenamefont {Shang},
  \citenamefont {Liu}, \citenamefont {Shao}, \citenamefont {Han}, \citenamefont
  {Zang}, \citenamefont {Zhang}, \citenamefont {Salama}, \citenamefont {Gao},
  \citenamefont {Lee}, \citenamefont {Thomale} \emph
  {et~al.}}]{shang2022experimental}%
  \BibitemOpen
  \bibfield  {author} {\bibinfo {author} {\bibfnamefont {C.}~\bibnamefont
  {Shang}}, \bibinfo {author} {\bibfnamefont {S.}~\bibnamefont {Liu}}, \bibinfo
  {author} {\bibfnamefont {R.}~\bibnamefont {Shao}}, \bibinfo {author}
  {\bibfnamefont {P.}~\bibnamefont {Han}}, \bibinfo {author} {\bibfnamefont
  {X.}~\bibnamefont {Zang}}, \bibinfo {author} {\bibfnamefont {X.}~\bibnamefont
  {Zhang}}, \bibinfo {author} {\bibfnamefont {K.~N.}\ \bibnamefont {Salama}},
  \bibinfo {author} {\bibfnamefont {W.}~\bibnamefont {Gao}}, \bibinfo {author}
  {\bibfnamefont {C.~H.}\ \bibnamefont {Lee}}, \bibinfo {author} {\bibfnamefont
  {R.}~\bibnamefont {Thomale}}, \emph {et~al.},\ }\href
  {https://doi.org/10.1002/advs.202202922} {\bibfield  {journal} {\bibinfo
  {journal} {Advanced Science}\ }\textbf {\bibinfo {volume} {9}},\ \bibinfo
  {pages} {2202922} (\bibinfo {year} {2022})}\BibitemShut {NoStop}%
\bibitem [{\citenamefont {Lin}\ \emph {et~al.}(2021)\citenamefont {Lin},
  \citenamefont {Ding}, \citenamefont {Ke},\ and\ \citenamefont
  {Li}}]{lin2021steering}%
  \BibitemOpen
  \bibfield  {author} {\bibinfo {author} {\bibfnamefont {Z.}~\bibnamefont
  {Lin}}, \bibinfo {author} {\bibfnamefont {L.}~\bibnamefont {Ding}}, \bibinfo
  {author} {\bibfnamefont {S.}~\bibnamefont {Ke}},\ and\ \bibinfo {author}
  {\bibfnamefont {X.}~\bibnamefont {Li}},\ }\href
  {https://doi.org/10.1364/OL.431904} {\bibfield  {journal} {\bibinfo
  {journal} {Optics Letters}\ }\textbf {\bibinfo {volume} {46}},\ \bibinfo
  {pages} {3512} (\bibinfo {year} {2021})}\BibitemShut {NoStop}%
\bibitem [{\citenamefont {Midya}\ \emph {et~al.}(2018)\citenamefont {Midya},
  \citenamefont {Zhao},\ and\ \citenamefont {Feng}}]{midya2018non}%
  \BibitemOpen
  \bibfield  {author} {\bibinfo {author} {\bibfnamefont {B.}~\bibnamefont
  {Midya}}, \bibinfo {author} {\bibfnamefont {H.}~\bibnamefont {Zhao}},\ and\
  \bibinfo {author} {\bibfnamefont {L.}~\bibnamefont {Feng}},\ }\href
  {https://doi.org/10.1038/s41467-018-05175-8} {\bibfield  {journal} {\bibinfo
  {journal} {Nature communications}\ }\textbf {\bibinfo {volume} {9}},\
  \bibinfo {pages} {2674} (\bibinfo {year} {2018})}\BibitemShut {NoStop}%
\bibitem [{\citenamefont {Longhi}(2017{\natexlab{a}})}]{longhi2017non}%
  \BibitemOpen
  \bibfield  {author} {\bibinfo {author} {\bibfnamefont {S.}~\bibnamefont
  {Longhi}},\ }\href {https://doi.org/10.1103/PhysRevB.95.014201} {\bibfield
  {journal} {\bibinfo  {journal} {Physical Review B}\ }\textbf {\bibinfo
  {volume} {95}},\ \bibinfo {pages} {014201} (\bibinfo {year}
  {2017}{\natexlab{a}})}\BibitemShut {NoStop}%
\bibitem [{\citenamefont {Longhi}\ \emph {et~al.}(2015)\citenamefont {Longhi},
  \citenamefont {Gatti},\ and\ \citenamefont {Della~Valle}}]{longhi2015non}%
  \BibitemOpen
  \bibfield  {author} {\bibinfo {author} {\bibfnamefont {S.}~\bibnamefont
  {Longhi}}, \bibinfo {author} {\bibfnamefont {D.}~\bibnamefont {Gatti}},\ and\
  \bibinfo {author} {\bibfnamefont {G.}~\bibnamefont {Della~Valle}},\ }\href
  {https://doi.org/10.1103/PhysRevB.92.094204} {\bibfield  {journal} {\bibinfo
  {journal} {Physical Review B}\ }\textbf {\bibinfo {volume} {92}},\ \bibinfo
  {pages} {094204} (\bibinfo {year} {2015})}\BibitemShut {NoStop}%
\bibitem [{\citenamefont
  {Longhi}(2017{\natexlab{b}})}]{longhi2017nonadiabatic}%
  \BibitemOpen
  \bibfield  {author} {\bibinfo {author} {\bibfnamefont {S.}~\bibnamefont
  {Longhi}},\ }\href {https://doi.org/10.1103/PhysRevA.95.062122} {\bibfield
  {journal} {\bibinfo  {journal} {Physical Review A}\ }\textbf {\bibinfo
  {volume} {95}},\ \bibinfo {pages} {062122} (\bibinfo {year}
  {2017}{\natexlab{b}})}\BibitemShut {NoStop}%
\bibitem [{\citenamefont {Wong}\ and\ \citenamefont
  {Oh}(2021)}]{wong2021topological}%
  \BibitemOpen
  \bibfield  {author} {\bibinfo {author} {\bibfnamefont {S.}~\bibnamefont
  {Wong}}\ and\ \bibinfo {author} {\bibfnamefont {S.~S.}\ \bibnamefont {Oh}},\
  }\href {https://doi.org/10.1103/PhysRevResearch.3.033042} {\bibfield
  {journal} {\bibinfo  {journal} {Physical Review Research}\ }\textbf {\bibinfo
  {volume} {3}},\ \bibinfo {pages} {033042} (\bibinfo {year}
  {2021})}\BibitemShut {NoStop}%
\bibitem [{\citenamefont {Chen}\ \emph {et~al.}(2019)\citenamefont {Chen},
  \citenamefont {Zhang}, \citenamefont {Xiong}, \citenamefont {Hang},
  \citenamefont {Li}, \citenamefont {Shen},\ and\ \citenamefont
  {Chan}}]{chen2019non}%
  \BibitemOpen
  \bibfield  {author} {\bibinfo {author} {\bibfnamefont {Y.}~\bibnamefont
  {Chen}}, \bibinfo {author} {\bibfnamefont {R.-Y.}\ \bibnamefont {Zhang}},
  \bibinfo {author} {\bibfnamefont {Z.}~\bibnamefont {Xiong}}, \bibinfo
  {author} {\bibfnamefont {Z.~H.}\ \bibnamefont {Hang}}, \bibinfo {author}
  {\bibfnamefont {J.}~\bibnamefont {Li}}, \bibinfo {author} {\bibfnamefont
  {J.~Q.}\ \bibnamefont {Shen}},\ and\ \bibinfo {author} {\bibfnamefont
  {C.~T.}\ \bibnamefont {Chan}},\ }\href
  {https://doi.org/10.1038/s41467-019-10974-8} {\bibfield  {journal} {\bibinfo
  {journal} {Nature communications}\ }\textbf {\bibinfo {volume} {10}},\
  \bibinfo {pages} {1} (\bibinfo {year} {2019})}\BibitemShut {NoStop}%
\bibitem [{\citenamefont {Yang}\ \emph {et~al.}(2019)\citenamefont {Yang},
  \citenamefont {Peng}, \citenamefont {Zhu}, \citenamefont {Buljan},
  \citenamefont {Joannopoulos}, \citenamefont {Zhen},\ and\ \citenamefont
  {Solja{\v{c}}i{\'c}}}]{yang2019synthesis}%
  \BibitemOpen
  \bibfield  {author} {\bibinfo {author} {\bibfnamefont {Y.}~\bibnamefont
  {Yang}}, \bibinfo {author} {\bibfnamefont {C.}~\bibnamefont {Peng}}, \bibinfo
  {author} {\bibfnamefont {D.}~\bibnamefont {Zhu}}, \bibinfo {author}
  {\bibfnamefont {H.}~\bibnamefont {Buljan}}, \bibinfo {author} {\bibfnamefont
  {J.~D.}\ \bibnamefont {Joannopoulos}}, \bibinfo {author} {\bibfnamefont
  {B.}~\bibnamefont {Zhen}},\ and\ \bibinfo {author} {\bibfnamefont
  {M.}~\bibnamefont {Solja{\v{c}}i{\'c}}},\ }\href
  {https://doi.org/10.1126/science.aay3183} {\bibfield  {journal} {\bibinfo
  {journal} {Science}\ }\textbf {\bibinfo {volume} {365}},\ \bibinfo {pages}
  {1021} (\bibinfo {year} {2019})}\BibitemShut {NoStop}%
\bibitem [{\citenamefont {Yang}\ \emph
  {et~al.}(2020{\natexlab{b}})\citenamefont {Yang}, \citenamefont {Yang},
  \citenamefont {You}, \citenamefont {Chan}, \citenamefont {Mao}, \citenamefont
  {Guo}, \citenamefont {Ma}, \citenamefont {Xia}, \citenamefont {Fan},
  \citenamefont {Xiang} \emph {et~al.}}]{yang2020observation}%
  \BibitemOpen
  \bibfield  {author} {\bibinfo {author} {\bibfnamefont {E.}~\bibnamefont
  {Yang}}, \bibinfo {author} {\bibfnamefont {B.}~\bibnamefont {Yang}}, \bibinfo
  {author} {\bibfnamefont {O.}~\bibnamefont {You}}, \bibinfo {author}
  {\bibfnamefont {H.-C.}\ \bibnamefont {Chan}}, \bibinfo {author}
  {\bibfnamefont {P.}~\bibnamefont {Mao}}, \bibinfo {author} {\bibfnamefont
  {Q.}~\bibnamefont {Guo}}, \bibinfo {author} {\bibfnamefont {S.}~\bibnamefont
  {Ma}}, \bibinfo {author} {\bibfnamefont {L.}~\bibnamefont {Xia}}, \bibinfo
  {author} {\bibfnamefont {D.}~\bibnamefont {Fan}}, \bibinfo {author}
  {\bibfnamefont {Y.}~\bibnamefont {Xiang}}, \emph {et~al.},\ }\href
  {https://doi.org/10.1103/PhysRevLett.125.033901} {\bibfield  {journal}
  {\bibinfo  {journal} {Physical Review Letters}\ }\textbf {\bibinfo {volume}
  {125}},\ \bibinfo {pages} {033901} (\bibinfo {year}
  {2020}{\natexlab{b}})}\BibitemShut {NoStop}%
\bibitem [{\citenamefont {Guo}\ \emph {et~al.}(2021)\citenamefont {Guo},
  \citenamefont {Jiang}, \citenamefont {Zhang}, \citenamefont {Zhang},
  \citenamefont {Zhang}, \citenamefont {Yang}, \citenamefont {Zhang},\ and\
  \citenamefont {Chan}}]{guo2021experimental}%
  \BibitemOpen
  \bibfield  {author} {\bibinfo {author} {\bibfnamefont {Q.}~\bibnamefont
  {Guo}}, \bibinfo {author} {\bibfnamefont {T.}~\bibnamefont {Jiang}}, \bibinfo
  {author} {\bibfnamefont {R.-Y.}\ \bibnamefont {Zhang}}, \bibinfo {author}
  {\bibfnamefont {L.}~\bibnamefont {Zhang}}, \bibinfo {author} {\bibfnamefont
  {Z.-Q.}\ \bibnamefont {Zhang}}, \bibinfo {author} {\bibfnamefont
  {B.}~\bibnamefont {Yang}}, \bibinfo {author} {\bibfnamefont {S.}~\bibnamefont
  {Zhang}},\ and\ \bibinfo {author} {\bibfnamefont {C.~T.}\ \bibnamefont
  {Chan}},\ }\href {https://doi.org/10.1038/s41586-021-03521-3} {\bibfield
  {journal} {\bibinfo  {journal} {Nature}\ }\textbf {\bibinfo {volume} {594}},\
  \bibinfo {pages} {195} (\bibinfo {year} {2021})}\BibitemShut {NoStop}%
\bibitem [{\citenamefont {Jiang}\ \emph {et~al.}(2021)\citenamefont {Jiang},
  \citenamefont {Bouhon}, \citenamefont {Lin}, \citenamefont {Zhou},
  \citenamefont {Hou}, \citenamefont {Li}, \citenamefont {Slager},\ and\
  \citenamefont {Jiang}}]{jiang2021experimental}%
  \BibitemOpen
  \bibfield  {author} {\bibinfo {author} {\bibfnamefont {B.}~\bibnamefont
  {Jiang}}, \bibinfo {author} {\bibfnamefont {A.}~\bibnamefont {Bouhon}},
  \bibinfo {author} {\bibfnamefont {Z.-K.}\ \bibnamefont {Lin}}, \bibinfo
  {author} {\bibfnamefont {X.}~\bibnamefont {Zhou}}, \bibinfo {author}
  {\bibfnamefont {B.}~\bibnamefont {Hou}}, \bibinfo {author} {\bibfnamefont
  {F.}~\bibnamefont {Li}}, \bibinfo {author} {\bibfnamefont {R.-J.}\
  \bibnamefont {Slager}},\ and\ \bibinfo {author} {\bibfnamefont {J.-H.}\
  \bibnamefont {Jiang}},\ }\href {https://doi.org/10.1038/s41567-021-01340-x}
  {\bibfield  {journal} {\bibinfo  {journal} {Nature Physics}\ }\textbf
  {\bibinfo {volume} {17}},\ \bibinfo {pages} {1239} (\bibinfo {year}
  {2021})}\BibitemShut {NoStop}%
\bibitem [{\citenamefont {Brosco}\ \emph {et~al.}(2021)\citenamefont {Brosco},
  \citenamefont {Pilozzi}, \citenamefont {Fazio},\ and\ \citenamefont
  {Conti}}]{brosco2021non}%
  \BibitemOpen
  \bibfield  {author} {\bibinfo {author} {\bibfnamefont {V.}~\bibnamefont
  {Brosco}}, \bibinfo {author} {\bibfnamefont {L.}~\bibnamefont {Pilozzi}},
  \bibinfo {author} {\bibfnamefont {R.}~\bibnamefont {Fazio}},\ and\ \bibinfo
  {author} {\bibfnamefont {C.}~\bibnamefont {Conti}},\ }\href
  {https://doi.org/10.1103/PhysRevA.103.063518} {\bibfield  {journal} {\bibinfo
   {journal} {Physical Review A}\ }\textbf {\bibinfo {volume} {103}},\ \bibinfo
  {pages} {063518} (\bibinfo {year} {2021})}\BibitemShut {NoStop}%
\bibitem [{\citenamefont {Chen}\ \emph {et~al.}(2022)\citenamefont {Chen},
  \citenamefont {Zhang}, \citenamefont {Chan},\ and\ \citenamefont
  {Ma}}]{chen2022classical}%
  \BibitemOpen
  \bibfield  {author} {\bibinfo {author} {\bibfnamefont {Z.-G.}\ \bibnamefont
  {Chen}}, \bibinfo {author} {\bibfnamefont {R.-Y.}\ \bibnamefont {Zhang}},
  \bibinfo {author} {\bibfnamefont {C.~T.}\ \bibnamefont {Chan}},\ and\
  \bibinfo {author} {\bibfnamefont {G.}~\bibnamefont {Ma}},\ }\href
  {https://doi.org/10.1038/s41567-021-01431-9} {\bibfield  {journal} {\bibinfo
  {journal} {Nature Physics}\ }\textbf {\bibinfo {volume} {18}},\ \bibinfo
  {pages} {179} (\bibinfo {year} {2022})}\BibitemShut {NoStop}%
\bibitem [{\citenamefont {Sun}\ \emph {et~al.}(2022)\citenamefont {Sun},
  \citenamefont {Zhang}, \citenamefont {Yu}, \citenamefont {Tian},
  \citenamefont {Chen},\ and\ \citenamefont {Sun}}]{sun2022non}%
  \BibitemOpen
  \bibfield  {author} {\bibinfo {author} {\bibfnamefont {Y.-K.}\ \bibnamefont
  {Sun}}, \bibinfo {author} {\bibfnamefont {X.-L.}\ \bibnamefont {Zhang}},
  \bibinfo {author} {\bibfnamefont {F.}~\bibnamefont {Yu}}, \bibinfo {author}
  {\bibfnamefont {Z.-N.}\ \bibnamefont {Tian}}, \bibinfo {author}
  {\bibfnamefont {Q.-D.}\ \bibnamefont {Chen}},\ and\ \bibinfo {author}
  {\bibfnamefont {H.-B.}\ \bibnamefont {Sun}},\ }\href
  {https://doi.org/10.1038/s41567-022-01669-x} {\bibfield  {journal} {\bibinfo
  {journal} {Nature Physics}\ }\textbf {\bibinfo {volume} {18}},\ \bibinfo
  {pages} {1080} (\bibinfo {year} {2022})}\BibitemShut {NoStop}%
\bibitem [{\citenamefont {You}\ \emph {et~al.}(2022)\citenamefont {You},
  \citenamefont {Liang}, \citenamefont {Xie}, \citenamefont {Gao},
  \citenamefont {Ye}, \citenamefont {Zhu},\ and\ \citenamefont
  {Zhang}}]{you2022observation}%
  \BibitemOpen
  \bibfield  {author} {\bibinfo {author} {\bibfnamefont {O.}~\bibnamefont
  {You}}, \bibinfo {author} {\bibfnamefont {S.}~\bibnamefont {Liang}}, \bibinfo
  {author} {\bibfnamefont {B.}~\bibnamefont {Xie}}, \bibinfo {author}
  {\bibfnamefont {W.}~\bibnamefont {Gao}}, \bibinfo {author} {\bibfnamefont
  {W.}~\bibnamefont {Ye}}, \bibinfo {author} {\bibfnamefont {J.}~\bibnamefont
  {Zhu}},\ and\ \bibinfo {author} {\bibfnamefont {S.}~\bibnamefont {Zhang}},\
  }\href {https://doi.org/10.1103/PhysRevLett.128.244302} {\bibfield  {journal}
  {\bibinfo  {journal} {Physical Review Letters}\ }\textbf {\bibinfo {volume}
  {128}},\ \bibinfo {pages} {244302} (\bibinfo {year} {2022})}\BibitemShut
  {NoStop}%
\bibitem [{\citenamefont {Noh}\ \emph {et~al.}(2020)\citenamefont {Noh},
  \citenamefont {Schuster}, \citenamefont {Iadecola}, \citenamefont {Huang},
  \citenamefont {Wang}, \citenamefont {Chen}, \citenamefont {Chamon},\ and\
  \citenamefont {Rechtsman}}]{noh2020braiding}%
  \BibitemOpen
  \bibfield  {author} {\bibinfo {author} {\bibfnamefont {J.}~\bibnamefont
  {Noh}}, \bibinfo {author} {\bibfnamefont {T.}~\bibnamefont {Schuster}},
  \bibinfo {author} {\bibfnamefont {T.}~\bibnamefont {Iadecola}}, \bibinfo
  {author} {\bibfnamefont {S.}~\bibnamefont {Huang}}, \bibinfo {author}
  {\bibfnamefont {M.}~\bibnamefont {Wang}}, \bibinfo {author} {\bibfnamefont
  {K.~P.}\ \bibnamefont {Chen}}, \bibinfo {author} {\bibfnamefont
  {C.}~\bibnamefont {Chamon}},\ and\ \bibinfo {author} {\bibfnamefont {M.~C.}\
  \bibnamefont {Rechtsman}},\ }\href
  {https://doi.org/10.1038/s41567-020-1007-5} {\bibfield  {journal} {\bibinfo
  {journal} {Nature Physics}\ }\textbf {\bibinfo {volume} {16}},\ \bibinfo
  {pages} {989} (\bibinfo {year} {2020})}\BibitemShut {NoStop}%
\bibitem [{\citenamefont {Zhang}\ \emph
  {et~al.}(2022{\natexlab{a}})\citenamefont {Zhang}, \citenamefont {Yu},
  \citenamefont {Chen}, \citenamefont {Tian}, \citenamefont {Chen},
  \citenamefont {Sun},\ and\ \citenamefont {Ma}}]{zhang2022non}%
  \BibitemOpen
  \bibfield  {author} {\bibinfo {author} {\bibfnamefont {X.-L.}\ \bibnamefont
  {Zhang}}, \bibinfo {author} {\bibfnamefont {F.}~\bibnamefont {Yu}}, \bibinfo
  {author} {\bibfnamefont {Z.-G.}\ \bibnamefont {Chen}}, \bibinfo {author}
  {\bibfnamefont {Z.-N.}\ \bibnamefont {Tian}}, \bibinfo {author}
  {\bibfnamefont {Q.-D.}\ \bibnamefont {Chen}}, \bibinfo {author}
  {\bibfnamefont {H.-B.}\ \bibnamefont {Sun}},\ and\ \bibinfo {author}
  {\bibfnamefont {G.}~\bibnamefont {Ma}},\ }\href
  {https://doi.org/10.1038/s41566-022-00976-2} {\bibfield  {journal} {\bibinfo
  {journal} {Nature Photonics}\ }\textbf {\bibinfo {volume} {16}},\ \bibinfo
  {pages} {390} (\bibinfo {year} {2022}{\natexlab{a}})}\BibitemShut {NoStop}%
\bibitem [{\citenamefont {Xu}\ \emph {et~al.}(2016)\citenamefont {Xu},
  \citenamefont {Sun}, \citenamefont {Han}, \citenamefont {Li}, \citenamefont
  {Pachos},\ and\ \citenamefont {Guo}}]{xu2016simulating}%
  \BibitemOpen
  \bibfield  {author} {\bibinfo {author} {\bibfnamefont {J.-S.}\ \bibnamefont
  {Xu}}, \bibinfo {author} {\bibfnamefont {K.}~\bibnamefont {Sun}}, \bibinfo
  {author} {\bibfnamefont {Y.-J.}\ \bibnamefont {Han}}, \bibinfo {author}
  {\bibfnamefont {C.-F.}\ \bibnamefont {Li}}, \bibinfo {author} {\bibfnamefont
  {J.~K.}\ \bibnamefont {Pachos}},\ and\ \bibinfo {author} {\bibfnamefont
  {G.-C.}\ \bibnamefont {Guo}},\ }\href {https://doi.org/10.1038/ncomms13194}
  {\bibfield  {journal} {\bibinfo  {journal} {Nature communications}\ }\textbf
  {\bibinfo {volume} {7}},\ \bibinfo {pages} {1} (\bibinfo {year}
  {2016})}\BibitemShut {NoStop}%
\bibitem [{\citenamefont {Yang}\ \emph
  {et~al.}(2020{\natexlab{c}})\citenamefont {Yang}, \citenamefont {Zhen},
  \citenamefont {Joannopoulos},\ and\ \citenamefont
  {Solja{\v{c}}i{\'{c}}}}]{yang20202d}%
  \BibitemOpen
  \bibfield  {author} {\bibinfo {author} {\bibfnamefont {Y.}~\bibnamefont
  {Yang}}, \bibinfo {author} {\bibfnamefont {B.}~\bibnamefont {Zhen}}, \bibinfo
  {author} {\bibfnamefont {J.~D.}\ \bibnamefont {Joannopoulos}},\ and\ \bibinfo
  {author} {\bibfnamefont {M.}~\bibnamefont {Solja{\v{c}}i{\'{c}}}},\ }\href
  {https://doi.org/10.1038/s41377-020-00384-7} {\bibfield  {journal} {\bibinfo
  {journal} {Light: Science {\&} Applications}\ }\textbf {\bibinfo {volume}
  {9}},\ \bibinfo {pages} {177} (\bibinfo {year}
  {2020}{\natexlab{c}})}\BibitemShut {NoStop}%
\bibitem [{\citenamefont {Yang}\ \emph {et~al.}(2022)\citenamefont {Yang},
  \citenamefont {Po}, \citenamefont {Liu}, \citenamefont {Joannopoulos},
  \citenamefont {Fu},\ and\ \citenamefont {Solja{\v{c}}i{\'c}}}]{yang2022non}%
  \BibitemOpen
  \bibfield  {author} {\bibinfo {author} {\bibfnamefont {Y.}~\bibnamefont
  {Yang}}, \bibinfo {author} {\bibfnamefont {H.~C.}\ \bibnamefont {Po}},
  \bibinfo {author} {\bibfnamefont {V.}~\bibnamefont {Liu}}, \bibinfo {author}
  {\bibfnamefont {J.~D.}\ \bibnamefont {Joannopoulos}}, \bibinfo {author}
  {\bibfnamefont {L.}~\bibnamefont {Fu}},\ and\ \bibinfo {author}
  {\bibfnamefont {M.}~\bibnamefont {Solja{\v{c}}i{\'c}}},\ }\href
  {https://doi.org/10.1103/PhysRevB.106.L161108} {\bibfield  {journal}
  {\bibinfo  {journal} {Physical Review B}\ }\textbf {\bibinfo {volume}
  {106}},\ \bibinfo {pages} {L161108} (\bibinfo {year} {2022})}\BibitemShut
  {NoStop}%
\bibitem [{\citenamefont {Gianfrate}\ \emph {et~al.}(2020)\citenamefont
  {Gianfrate}, \citenamefont {Bleu}, \citenamefont {Dominici}, \citenamefont
  {Ardizzone}, \citenamefont {De~Giorgi}, \citenamefont {Ballarini},
  \citenamefont {Lerario}, \citenamefont {West}, \citenamefont {Pfeiffer},
  \citenamefont {Solnyshkov}, \citenamefont {Sanvitto},\ and\ \citenamefont
  {Malpuech}}]{gianfrate2020direct}%
  \BibitemOpen
  \bibfield  {author} {\bibinfo {author} {\bibfnamefont {A.}~\bibnamefont
  {Gianfrate}}, \bibinfo {author} {\bibfnamefont {O.}~\bibnamefont {Bleu}},
  \bibinfo {author} {\bibfnamefont {L.}~\bibnamefont {Dominici}}, \bibinfo
  {author} {\bibfnamefont {V.}~\bibnamefont {Ardizzone}}, \bibinfo {author}
  {\bibfnamefont {M.}~\bibnamefont {De~Giorgi}}, \bibinfo {author}
  {\bibfnamefont {D.}~\bibnamefont {Ballarini}}, \bibinfo {author}
  {\bibfnamefont {G.}~\bibnamefont {Lerario}}, \bibinfo {author} {\bibfnamefont
  {K.~W.}\ \bibnamefont {West}}, \bibinfo {author} {\bibfnamefont {L.~N.}\
  \bibnamefont {Pfeiffer}}, \bibinfo {author} {\bibfnamefont {D.~D.}\
  \bibnamefont {Solnyshkov}}, \bibinfo {author} {\bibfnamefont
  {D.}~\bibnamefont {Sanvitto}},\ and\ \bibinfo {author} {\bibfnamefont
  {G.}~\bibnamefont {Malpuech}},\ }\href
  {https://doi.org/10.1038/s41586-020-1989-2} {\bibfield  {journal} {\bibinfo
  {journal} {Nature}\ }\textbf {\bibinfo {volume} {578}},\ \bibinfo {pages}
  {381} (\bibinfo {year} {2020})}\BibitemShut {NoStop}%
\bibitem [{\citenamefont {Whittaker}\ \emph {et~al.}(2021)\citenamefont
  {Whittaker}, \citenamefont {Dowling}, \citenamefont {Nalitov}, \citenamefont
  {Yulin}, \citenamefont {Royall}, \citenamefont {Clarke}, \citenamefont
  {Skolnick}, \citenamefont {Shelykh},\ and\ \citenamefont
  {Krizhanovskii}}]{whittaker2021optical}%
  \BibitemOpen
  \bibfield  {author} {\bibinfo {author} {\bibfnamefont {C.}~\bibnamefont
  {Whittaker}}, \bibinfo {author} {\bibfnamefont {T.}~\bibnamefont {Dowling}},
  \bibinfo {author} {\bibfnamefont {A.}~\bibnamefont {Nalitov}}, \bibinfo
  {author} {\bibfnamefont {A.}~\bibnamefont {Yulin}}, \bibinfo {author}
  {\bibfnamefont {B.}~\bibnamefont {Royall}}, \bibinfo {author} {\bibfnamefont
  {E.}~\bibnamefont {Clarke}}, \bibinfo {author} {\bibfnamefont
  {M.}~\bibnamefont {Skolnick}}, \bibinfo {author} {\bibfnamefont
  {I.}~\bibnamefont {Shelykh}},\ and\ \bibinfo {author} {\bibfnamefont
  {D.}~\bibnamefont {Krizhanovskii}},\ }\href
  {https://doi.org/10.1038/s41566-020-00729-z} {\bibfield  {journal} {\bibinfo
  {journal} {Nature Photonics}\ }\textbf {\bibinfo {volume} {15}},\ \bibinfo
  {pages} {193} (\bibinfo {year} {2021})}\BibitemShut {NoStop}%
\bibitem [{\citenamefont {Iadecola}\ \emph {et~al.}(2016)\citenamefont
  {Iadecola}, \citenamefont {Schuster},\ and\ \citenamefont
  {Chamon}}]{iadecola2016non}%
  \BibitemOpen
  \bibfield  {author} {\bibinfo {author} {\bibfnamefont {T.}~\bibnamefont
  {Iadecola}}, \bibinfo {author} {\bibfnamefont {T.}~\bibnamefont {Schuster}},\
  and\ \bibinfo {author} {\bibfnamefont {C.}~\bibnamefont {Chamon}},\ }\href
  {https://doi.org/10.1103/PhysRevLett.117.073901} {\bibfield  {journal}
  {\bibinfo  {journal} {Phys. Rev. Lett.}\ }\textbf {\bibinfo {volume} {117}},\
  \bibinfo {pages} {073901} (\bibinfo {year} {2016})}\BibitemShut {NoStop}%
\bibitem [{\citenamefont {Polimeno}\ \emph {et~al.}(2021)\citenamefont
  {Polimeno}, \citenamefont {Fieramosca}, \citenamefont {Lerario},
  \citenamefont {De~Marco}, \citenamefont {De~Giorgi}, \citenamefont
  {Ballarini}, \citenamefont {Dominici}, \citenamefont {Ardizzone},
  \citenamefont {Pugliese}, \citenamefont {Prontera} \emph
  {et~al.}}]{polimeno2021experimental}%
  \BibitemOpen
  \bibfield  {author} {\bibinfo {author} {\bibfnamefont {L.}~\bibnamefont
  {Polimeno}}, \bibinfo {author} {\bibfnamefont {A.}~\bibnamefont
  {Fieramosca}}, \bibinfo {author} {\bibfnamefont {G.}~\bibnamefont {Lerario}},
  \bibinfo {author} {\bibfnamefont {L.}~\bibnamefont {De~Marco}}, \bibinfo
  {author} {\bibfnamefont {M.}~\bibnamefont {De~Giorgi}}, \bibinfo {author}
  {\bibfnamefont {D.}~\bibnamefont {Ballarini}}, \bibinfo {author}
  {\bibfnamefont {L.}~\bibnamefont {Dominici}}, \bibinfo {author}
  {\bibfnamefont {V.}~\bibnamefont {Ardizzone}}, \bibinfo {author}
  {\bibfnamefont {M.}~\bibnamefont {Pugliese}}, \bibinfo {author}
  {\bibfnamefont {C.}~\bibnamefont {Prontera}}, \emph {et~al.},\ }\href
  {https://doi.org/10.1364/OPTICA.427088} {\bibfield  {journal} {\bibinfo
  {journal} {Optica}\ }\textbf {\bibinfo {volume} {8}},\ \bibinfo {pages}
  {1442} (\bibinfo {year} {2021})}\BibitemShut {NoStop}%
\bibitem [{\citenamefont {Cheng}\ \emph {et~al.}(2023)\citenamefont {Cheng},
  \citenamefont {Wang},\ and\ \citenamefont {Fan}}]{cheng2023artificial}%
  \BibitemOpen
  \bibfield  {author} {\bibinfo {author} {\bibfnamefont {D.}~\bibnamefont
  {Cheng}}, \bibinfo {author} {\bibfnamefont {K.}~\bibnamefont {Wang}},\ and\
  \bibinfo {author} {\bibfnamefont {S.}~\bibnamefont {Fan}},\ }\href
  {https://doi.org/10.1103/PhysRevLett.130.083601} {\bibfield  {journal}
  {\bibinfo  {journal} {Physical Review Letters}\ }\textbf {\bibinfo {volume}
  {130}},\ \bibinfo {pages} {083601} (\bibinfo {year} {2023})}\BibitemShut
  {NoStop}%
\bibitem [{\citenamefont {Aidelsburger}\ \emph {et~al.}(2018)\citenamefont
  {Aidelsburger}, \citenamefont {Nascimbene},\ and\ \citenamefont
  {Goldman}}]{aidelsburger2018artificial}%
  \BibitemOpen
  \bibfield  {author} {\bibinfo {author} {\bibfnamefont {M.}~\bibnamefont
  {Aidelsburger}}, \bibinfo {author} {\bibfnamefont {S.}~\bibnamefont
  {Nascimbene}},\ and\ \bibinfo {author} {\bibfnamefont {N.}~\bibnamefont
  {Goldman}},\ }\href {https://doi.org/10.1016/j.crhy.2018.03.002} {\bibfield
  {journal} {\bibinfo  {journal} {Comptes Rendus Physique}\ }\textbf {\bibinfo
  {volume} {19}},\ \bibinfo {pages} {394} (\bibinfo {year} {2018})}\BibitemShut
  {NoStop}%
\bibitem [{\citenamefont {Wu}\ \emph {et~al.}(2022)\citenamefont {Wu},
  \citenamefont {Wang}, \citenamefont {Biao}, \citenamefont {Fei},
  \citenamefont {Zhang}, \citenamefont {Yin}, \citenamefont {Hu}, \citenamefont
  {Song}, \citenamefont {Wu}, \citenamefont {Song} \emph {et~al.}}]{wu2022non}%
  \BibitemOpen
  \bibfield  {author} {\bibinfo {author} {\bibfnamefont {J.}~\bibnamefont
  {Wu}}, \bibinfo {author} {\bibfnamefont {Z.}~\bibnamefont {Wang}}, \bibinfo
  {author} {\bibfnamefont {Y.}~\bibnamefont {Biao}}, \bibinfo {author}
  {\bibfnamefont {F.}~\bibnamefont {Fei}}, \bibinfo {author} {\bibfnamefont
  {S.}~\bibnamefont {Zhang}}, \bibinfo {author} {\bibfnamefont
  {Z.}~\bibnamefont {Yin}}, \bibinfo {author} {\bibfnamefont {Y.}~\bibnamefont
  {Hu}}, \bibinfo {author} {\bibfnamefont {Z.}~\bibnamefont {Song}}, \bibinfo
  {author} {\bibfnamefont {T.}~\bibnamefont {Wu}}, \bibinfo {author}
  {\bibfnamefont {F.}~\bibnamefont {Song}}, \emph {et~al.},\ }\href
  {https://doi.org/10.1038/s41928-022-00833-8} {\bibfield  {journal} {\bibinfo
  {journal} {Nature Electronics}\ }\textbf {\bibinfo {volume} {5}},\ \bibinfo
  {pages} {635} (\bibinfo {year} {2022})}\BibitemShut {NoStop}%
\bibitem [{\citenamefont {Lin}\ \emph {et~al.}(2011)\citenamefont {Lin},
  \citenamefont {Jim{\'e}nez-Garc{\'\i}a},\ and\ \citenamefont
  {Spielman}}]{lin2011spin}%
  \BibitemOpen
  \bibfield  {author} {\bibinfo {author} {\bibfnamefont {Y.-J.}\ \bibnamefont
  {Lin}}, \bibinfo {author} {\bibfnamefont {K.}~\bibnamefont
  {Jim{\'e}nez-Garc{\'\i}a}},\ and\ \bibinfo {author} {\bibfnamefont {I.~B.}\
  \bibnamefont {Spielman}},\ }\href {https://doi.org/10.1038/nature09887}
  {\bibfield  {journal} {\bibinfo  {journal} {Nature}\ }\textbf {\bibinfo
  {volume} {471}},\ \bibinfo {pages} {83} (\bibinfo {year} {2011})}\BibitemShut
  {NoStop}%
\bibitem [{\citenamefont {Hatano}\ and\ \citenamefont
  {Nelson}(1996)}]{hatano1996localization}%
  \BibitemOpen
  \bibfield  {author} {\bibinfo {author} {\bibfnamefont {N.}~\bibnamefont
  {Hatano}}\ and\ \bibinfo {author} {\bibfnamefont {D.~R.}\ \bibnamefont
  {Nelson}},\ }\href {https://doi.org/10.1103/PhysRevLett.77.570} {\bibfield
  {journal} {\bibinfo  {journal} {Physical review letters}\ }\textbf {\bibinfo
  {volume} {77}},\ \bibinfo {pages} {570} (\bibinfo {year} {1996})}\BibitemShut
  {NoStop}%
\bibitem [{\citenamefont {Zhou}(2023)}]{zhou2023non}%
  \BibitemOpen
  \bibfield  {author} {\bibinfo {author} {\bibfnamefont {L.}~\bibnamefont
  {Zhou}},\ }\bibfield  {journal} {\bibinfo  {journal} {arXiv preprint
  arXiv:2302.05710}\ }\href {https://doi.org/10.48550/arXiv.2302.05710}
  {10.48550/arXiv.2302.05710} (\bibinfo {year} {2023})\BibitemShut {NoStop}%
\bibitem [{\citenamefont {Yang}\ and\ \citenamefont {Hu}(2019)}]{yang2019non}%
  \BibitemOpen
  \bibfield  {author} {\bibinfo {author} {\bibfnamefont {Z.}~\bibnamefont
  {Yang}}\ and\ \bibinfo {author} {\bibfnamefont {J.}~\bibnamefont {Hu}},\
  }\href {https://doi.org/10.1103/PhysRevB.99.081102} {\bibfield  {journal}
  {\bibinfo  {journal} {Physical Review B}\ }\textbf {\bibinfo {volume} {99}},\
  \bibinfo {pages} {081102} (\bibinfo {year} {2019})}\BibitemShut {NoStop}%
\bibitem [{\citenamefont {Carlstr{\"o}m}\ and\ \citenamefont
  {Bergholtz}(2018)}]{carlstrom2018exceptional}%
  \BibitemOpen
  \bibfield  {author} {\bibinfo {author} {\bibfnamefont {J.}~\bibnamefont
  {Carlstr{\"o}m}}\ and\ \bibinfo {author} {\bibfnamefont {E.~J.}\ \bibnamefont
  {Bergholtz}},\ }\href {https://doi.org/10.1103/PhysRevA.98.042114} {\bibfield
   {journal} {\bibinfo  {journal} {Physical Review A}\ }\textbf {\bibinfo
  {volume} {98}},\ \bibinfo {pages} {042114} (\bibinfo {year}
  {2018})}\BibitemShut {NoStop}%
\bibitem [{\citenamefont {Zhang}\ \emph
  {et~al.}(2022{\natexlab{b}})\citenamefont {Zhang}, \citenamefont {Zhang},
  \citenamefont {Lu},\ and\ \citenamefont {Chen}}]{zhang2022review}%
  \BibitemOpen
  \bibfield  {author} {\bibinfo {author} {\bibfnamefont {X.}~\bibnamefont
  {Zhang}}, \bibinfo {author} {\bibfnamefont {T.}~\bibnamefont {Zhang}},
  \bibinfo {author} {\bibfnamefont {M.-H.}\ \bibnamefont {Lu}},\ and\ \bibinfo
  {author} {\bibfnamefont {Y.-F.}\ \bibnamefont {Chen}},\ }\href
  {https://doi.org/10.1080/23746149.2022.2109431} {\bibfield  {journal}
  {\bibinfo  {journal} {Advances in Physics: X}\ }\textbf {\bibinfo {volume}
  {7}},\ \bibinfo {pages} {2109431} (\bibinfo {year}
  {2022}{\natexlab{b}})}\BibitemShut {NoStop}%
\bibitem [{\citenamefont {Ashida}\ \emph {et~al.}(2020)\citenamefont {Ashida},
  \citenamefont {Gong},\ and\ \citenamefont {Ueda}}]{ashida2020non}%
  \BibitemOpen
  \bibfield  {author} {\bibinfo {author} {\bibfnamefont {Y.}~\bibnamefont
  {Ashida}}, \bibinfo {author} {\bibfnamefont {Z.}~\bibnamefont {Gong}},\ and\
  \bibinfo {author} {\bibfnamefont {M.}~\bibnamefont {Ueda}},\ }\href
  {https://doi.org/10.1080/00018732.2021.1876991} {\bibfield  {journal}
  {\bibinfo  {journal} {Advances in Physics}\ }\textbf {\bibinfo {volume}
  {69}},\ \bibinfo {pages} {249} (\bibinfo {year} {2020})}\BibitemShut
  {NoStop}%
\bibitem [{SM_()}]{SM_note}%
  \BibitemOpen
  \href@noop {} {}\bibinfo {note} {See Supplementary Materials for additional
  notes on (i) Hopf link, (ii) Non-Bloch wave approach, (iii) Winding-number
  approach, (iv) zero modes, and (v) asymptotic analysis.}\BibitemShut {Stop}%
\bibitem [{\citenamefont {Dutt}\ \emph {et~al.}(2019)\citenamefont {Dutt},
  \citenamefont {Minkov}, \citenamefont {Lin}, \citenamefont {Yuan},
  \citenamefont {Miller},\ and\ \citenamefont {Fan}}]{dutt2019experimental}%
  \BibitemOpen
  \bibfield  {author} {\bibinfo {author} {\bibfnamefont {A.}~\bibnamefont
  {Dutt}}, \bibinfo {author} {\bibfnamefont {M.}~\bibnamefont {Minkov}},
  \bibinfo {author} {\bibfnamefont {Q.}~\bibnamefont {Lin}}, \bibinfo {author}
  {\bibfnamefont {L.}~\bibnamefont {Yuan}}, \bibinfo {author} {\bibfnamefont
  {D.~A.}\ \bibnamefont {Miller}},\ and\ \bibinfo {author} {\bibfnamefont
  {S.}~\bibnamefont {Fan}},\ }\href
  {https://doi.org/10.1038/s41467-019-11117-9} {\bibfield  {journal} {\bibinfo
  {journal} {Nature communications}\ }\textbf {\bibinfo {volume} {10}},\
  \bibinfo {pages} {3122} (\bibinfo {year} {2019})}\BibitemShut {NoStop}%
\bibitem [{\citenamefont {Regensburger}\ \emph {et~al.}(2011)\citenamefont
  {Regensburger}, \citenamefont {Bersch}, \citenamefont {Hinrichs},
  \citenamefont {Onishchukov}, \citenamefont {Schreiber}, \citenamefont
  {Silberhorn},\ and\ \citenamefont {Peschel}}]{regensburger2011photon}%
  \BibitemOpen
  \bibfield  {author} {\bibinfo {author} {\bibfnamefont {A.}~\bibnamefont
  {Regensburger}}, \bibinfo {author} {\bibfnamefont {C.}~\bibnamefont
  {Bersch}}, \bibinfo {author} {\bibfnamefont {B.}~\bibnamefont {Hinrichs}},
  \bibinfo {author} {\bibfnamefont {G.}~\bibnamefont {Onishchukov}}, \bibinfo
  {author} {\bibfnamefont {A.}~\bibnamefont {Schreiber}}, \bibinfo {author}
  {\bibfnamefont {C.}~\bibnamefont {Silberhorn}},\ and\ \bibinfo {author}
  {\bibfnamefont {U.}~\bibnamefont {Peschel}},\ }\href
  {https://doi.org/10.1103/PhysRevLett.107.233902} {\bibfield  {journal}
  {\bibinfo  {journal} {Physical review letters}\ }\textbf {\bibinfo {volume}
  {107}},\ \bibinfo {pages} {233902} (\bibinfo {year} {2011})}\BibitemShut
  {NoStop}%
\end{thebibliography}
\end{document}